\documentstyle[newlfont,epsf,macros,newcomnd,12pt,din_a4
]{article}
\newcommand{\two}[2]{#2}   
\newlength{\pcm}
\newlength{\pmm}
\newcommand {\GB} {\,\epsfxsize=1.2\pcm \parbox{1.2\pcm}{\epsfbox{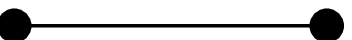}}\,}
\newcommand {\GBdotted} {\,\epsfxsize=1.2\pcm \parbox{1.2\pcm}{\epsfbox{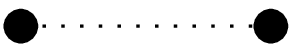}}\,}
\newcommand {\GM} {\,\epsfxsize=1.5\pcm \parbox{1.5\pcm}{\epsfbox{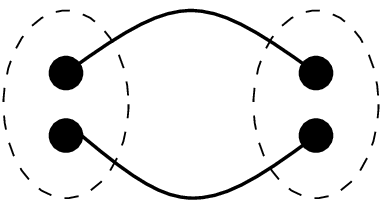}}\,}
\newcommand {\GH} {\,\epsfxsize=0.8\pcm \parbox{0.8\pcm}{\epsfbox{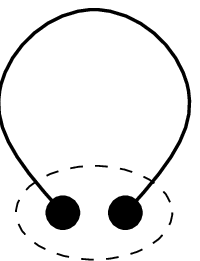}}\,}
\newcommand {\GO} {\,\epsfxsize=0.4\pcm \parbox{0.4\pcm}{\epsfbox{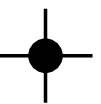}}\,}
\newcommand {\FD} {\,\epsfxsize=1.7\pcm \parbox{1.7\pcm}{\epsfbox{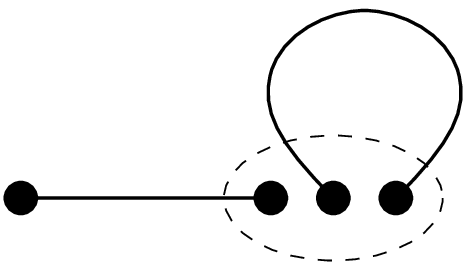}}\,}

\newcommand{\Res}{\mbox{Res}}

\newlength{\ul}
\newcommand{\PA}{\!\!\!\raisebox{-0.4\ul}
{\setlength{\unitlength}{0.1\ul}
\begin{picture}(11,10)(0,0)
\thicklines
\put(1,5){\put(0,0){\circle*2}
\put(5,0){\circle{10}}
}\end{picture}}}
\newcommand{\PB}{\!\!\!\raisebox{-0.4\ul}
{\setlength{\unitlength}{0.1\ul}
\begin{picture}(26,10)(0,0)
\thicklines
\put(16,5){
\put(-15,0){\circle*2}
\put(-5,0){\dashbox{0.5}(5,0){}}
\put(-5,0){\circle*2}
\put(0,0){\circle*2}
\put(5,0){\circle{10}}
\put(-10,0){\circle{10}}
}\end{picture}}}
\newcommand{\PC}{\!\!\!\raisebox{-0.4\ul}
{\setlength{\unitlength}{0.1\ul}
\begin{picture}(30,10)(0,0)
\thicklines
\put(10,5){
\put(-5,0){\dashbox{0.5}(5,0){}}
\put(-5,0){\circle*2}
\put(0,0){\circle*2}
\put(-10,0){\oval(10,10)[r]}
\put(5,0){\circle{10}}
\put(10,0){\dashbox{0.5}(5,0){}}
\put(20,0){\oval(10,10)[l]}
\put(10,0){\circle*2}
\put(15,0){\circle*2}
}\end{picture}}}
\newcommand{\PCclose}{\!\!\!\raisebox{-0.4\ul}
{\setlength{\unitlength}{0.1\ul}
\begin{picture}(40,10)(-5,0)
\thicklines
\put(10,5){
\put(-5,0){\dashbox{0.5}(5,0){}}
\put(-5,0){\circle*2}
\put(0,0){\circle*2}
\put(-10,0){\circle{10}}
\put(5,0){\circle{10}}
\put(10,0){\dashbox{0.5}(5,0){}}
\put(20,0){\circle{10}}
\put(10,0){\circle*2}
\put(15,0){\circle*2}
}\end{picture}}}
\newcommand{\PD}{\!\!\!\raisebox{-0.4\ul}
{\setlength{\unitlength}{0.1\ul}
\begin{picture}(15,10)(0,0)
\thicklines
\put(5,5){
\put(10,0){\oval(10,10)[l]}
\put(-5,0){\oval(10,10)[r]}
\put(0,0){\circle*2}
\put(0,0){\circle*2}
\put(0,0){\dashbox{0.5}(5,0){}}
\put(5,0){\circle*2}
}\end{picture}}}
\newcommand{\PE}{{
\!\!\!\raisebox{-0.65\ul}
{\setlength{\unitlength}{0.1\ul}
\begin{picture}(15,15)(0,0)
\thicklines
\put(5,5){
\put(10,2.5){\oval(10,15)[l]}
\put(-5,2.5){\oval(10,15)[r]}
\put(0,0){\circle*2}
\put(0,5){\circle*2}
\put(0,0){\circle*2}
\put(0,0){\dashbox{0.5}(5,0){}}
\put(0,5){\dashbox{0.5}(5,0){}}
\put(5,0){\circle*2}
\put(5,5){\circle*2}
}\end{picture}}}}
\newcommand{\PEclose}{{
\!\!\!\raisebox{-0.65\ul}
{\setlength{\unitlength}{0.1\ul}
\begin{picture}(25,15)(-5,0)
\thicklines
\put(5,5){
\put(10,2.5){\oval(10,15)[l]}
\put(10,2.5){\oval(10,15)[r]}
\put(-5,2.5){\oval(10,15)[r]}
\put(-5,2.5){\oval(10,15)[l]}
\put(0,0){\circle*2}
\put(0,5){\circle*2}
\put(0,0){\circle*2}
\put(0,0){\dashbox{0.5}(5,0){}}
\put(0,5){\dashbox{0.5}(5,0){}}
\put(5,0){\circle*2}
\put(5,5){\circle*2}
}\end{picture}}}}
\newcommand{\PFclose}{\!\!\!\raisebox{-0.4\ul}
{\setlength{\unitlength}{0.1\ul}
\begin{picture}(35,12)(-5,0)
\thicklines
\put(10,5){\put(0,0){\circle*2}
\put(-5,5){\circle*{2}}
\put(-5,-5){\circle*{2}}
\put(-10,0){\oval(20,10)[r]}
\put(-10,0){\oval(10,10)[l]}
\put(-5,-5){\dashbox{0.5}(0,10){}}
\put(0,0){\dashbox{0.5}(5,0){}}
\put(5,0){\put(0,0){\circle*2}}
\put(15,0){\oval(20,10)[l]}
\put(15,0){\oval(10,10)[r]}
}\end{picture}}}\newcommand{\PF}{\!\!\!\raisebox{-0.4\ul}
{\setlength{\unitlength}{0.1\ul}
\begin{picture}(25,12)(0,0)
\thicklines
\put(10,5){\put(0,0){\circle*2}
\put(-5,5){\circle*{2}}
\put(-5,-5){\circle*{2}}
\put(-10,0){\oval(20,10)[r]}
\put(-5,-5){\dashbox{0.5}(0,10){}}
\put(0,0){\dashbox{0.5}(5,0){}}
\put(5,0){\put(0,0){\circle*2}}
\put(15,0){\oval(20,10)[l]}
}\end{picture}}}
\newcommand{\PG}{\!\!\!\raisebox{-0.4\ul}
{\setlength{\unitlength}{0.1\ul}
\begin{picture}(16,10)(0,0)
\thicklines
\put(1,5){
	\put(0,0){\circle*2}
	\put(7.5,5){\circle*{2}}
	\put(7.5,-5){\circle*{2}}
	\put(7.5,0){\oval(15,10)}
	\put(7.5,-5){\dashbox{0.5}(0,10){}}
}\end{picture}}}
\newcommand{\PH}{\!\!\!\raisebox{-0.4\ul}
{\setlength{\unitlength}{0.1\ul}
\begin{picture}(12,10)(0,0)
\thicklines
\put(1,5){
	\put(0,0){\circle*2}
	\put(5,0){\circle{10}}
	\put(10,0){\circle*2}
}\end{picture}}}
\newcommand{\PI}{\!\!\!\raisebox{-0.4\ul}
{\setlength{\unitlength}{0.1\ul}
\begin{picture}(21,10)(0,0)
\thicklines
\put(16,5){
\put(-15,0){\circle*2}
\put(-5,0){\circle*2}
\put(0,0){\circle{10}}
\put(-10,0){\circle{10}}
}\end{picture}}}
\newcommand{\PJ}{\!\!\!\raisebox{-0.4\ul}
{\setlength{\unitlength}{0.1\ul}
\begin{picture}(42,10)(-7,0)
\thicklines
\put(10,5){
\put(-5,0){\dashbox{0.5}(5,0){}}
\put(-15,0){\circle*2}
\put(-5,0){\circle*2}
\put(0,0){\circle*2}
\put(-10,0){\circle{10}}
\put(5,0){\circle{10}}
\put(10,0){\dashbox{0.5}(5,0){}}
\put(20,0){\circle{10}}
\put(10,0){\circle*2}
\put(15,0){\circle*2}
}\end{picture}}}
\newcommand{\PK}{\!\!\!\raisebox{-0.4\ul}
{\setlength{\unitlength}{0.1\ul}
\begin{picture}(27,25)(-7,0)
\thicklines
\put(10,5){
\put(-5,0){\dashbox{0.5}(5,0){}}
\put(-15,0){\circle*2}
\put(-5,0){\circle*2}
\put(0,0){\circle*2}
\put(-10,0){\circle{10}}
\put(5,0){\circle{10}}
\put(-10,5){\dashbox{0.5}(0,5){}}
\put(-10,15){\circle{10}}
\put(-10,5){\circle*2}
\put(-10,10){\circle*2}
}\end{picture}}}

\begin{document}

\begin{titlepage}

\noindent
\renewcommand{\thefootnote}{\fnsymbol{footnote}}
\parbox{1.85cm}{\epsfxsize=1.85cm \epsfbox{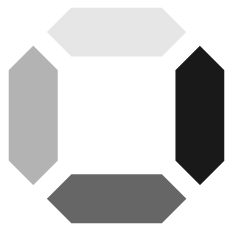}}\hfill%
\begin{minipage}{10cm}
\rightline{Uni GH Essen- and MIT-preprint}%
\rightline{cond-mat/9803389}%
\rightline{\today}%
\end{minipage}
\vskip 1.3cm
\centerline{\bf\sf\Large Generalizing the $O(N)$-field theory  to 
$N$-colored}
\smallskip
\centerline{\bf\sf\Large  manifolds of arbitrary internal dimension $D$}
\vfill
\centerline{\bf\large Kay J\"org Wiese%
\footnote{Email: wiese@next23.theo-phys.uni-essen.de}}
\smallskip
\centerline{\small Fachbereich Physik, Universit\"at GH Essen,  45117 Essen,
Germany}
\smallskip\smallskip
\smallskip\smallskip
\centerline{\bf\large Mehran Kardar%
\footnote{Email: kardar@mit.edu}}\smallskip
\centerline{\small Department of Physics, MIT, Cambridge, Massachusetts 02139, USA}

\vfill
\vspace{-5mm}
\begin{abstract}
We introduce a geometric generalization of the $O(N)$-field theory that describes 
$N$-colored membranes with arbitrary dimension $D$. 
As the $O(N)$-model reduces in the limit $N\to0$ to  
self-avoiding polymers, the $N$-colored manifold model leads to self-avoiding
tethered membranes. In the other limit, for inner dimension $D\to1$, the manifold model
reduces to the $O(N)$-field theory. 
We analyze the scaling properties of the model at criticality by a one-loop 
perturbative renormalization group analysis around an upper critical line.
The freedom to optimize with respect to the expansion point on this line
allows us to obtain the exponent $\nu$ of standard field theory to
much better precision that the  usual 1-loop calculations. 
Some other field theoretical techniques, such as the large $N$ limit and
Hartree approximation, can also be applied to this model.
By comparison of low and high temperature expansions,
we arrive at a conjecture for the nature of droplets dominating the $3d$-Ising model 
at criticality, which is satisfied by our numerical results. 
We can also construct an appropriate generalization that describes
cubic anisotropy, by adding an interaction between manifolds of the same color.
The two parameter space includes a variety of new phases and fixed points, 
some with Ising criticality, enabling us to extract a remarkably precise
value of 0.6315 for the exponent $\nu$ in $d=3$. 
A particular limit of the model with cubic anisotropy corresponds to the
random bond Ising problem; unlike the field theory formulation, we find
a fixed point describing this system at 1-loop order.

\end{abstract}
\vspace{-5mm}
\vfill

\centerline{\em Submitted for publication in Nuclear Physics B}

\vfill

\end{titlepage}\addtocounter{page}{1}
\renewcommand{\thefootnote}{\fnsymbol{footnote}}

\two{\begin{multicols}{2}\narrowtext}{}
\tableofcontents \newpage
\noindent
\section{Introduction}

 Field theoretical models are particularly suited for description of
{\em universal quantities} which do not depend on the details of the system.
This is epitomized by systems undergoing  symmetry breaking 
continuous phase transitions. Their critical behavior is described
by a set of exponents which are completely characterized
by dimension and the underlying symmetry (the number of 
components of the  order parameter). 
Universality is assured since the microscopic details are averaged
out, and do not effect the large scale fluctuations  (for a review, see 
Ref.~\cite{KogutWilson}).
The universal features are thus captured by the  $O(N)$ model,
which is a field theory for the statistics of  $N$-component spins. 
A variety of techniques have been developed to obtain the critical
behavior of this model; possibly the most successful is the
renormalization group procedure\cite{KogutWilson} which analytically
justifies the concept of universality. 
The most technically convenient implementations are field theoretical methods,
e.g.\ the $\E$-expansion about the upper critical dimension of 4, 
an  expansion about the lower critical dimension of 2,
and exact resummations in the large $N$ limit.
(For an overview of these techniques, see  Ref.\cite{Zinn}.)
The best studied method is the $\E$-expansion about the upper
critical dimension 4, where calculations have been performed up 
to  5-loop order. Together with resummation techniques which take
care of the large-order behavior known from instanton calculus,
this is a very powerful tool for extracting critical exponents.

On the other hand, field theories have strong connections to 
geometrical problems involving fluctuating lines.
 For example, the motion of particles in space-time describes a world-line.
Summing over all world-lines, weighted by an appropriate action, is the Feynman
path integral approach to calculating transition probabilities, 
which can alternatively be obtained from a quantum field theory.
Another example  is the high-temperature expansion of the Ising model. 
The energy-energy correlation function can be expressed as a  sum over all 
self-avoiding closed loops which pass through two given points.
The generalization to $N$-component spins is straightforward:
The partition function of the corresponding `loop model' is obtained by summing
over all configurations of a gas of closed loops, where each loop comes in 
$N$ colors, or has a fugacity of $N$.
In the limit $N \to 0$, only a single loop contributes, giving the partition function 
of a closed self-avoiding polymer.
 For $N>0$  the model describes polymers which can 
break up and polymerize dynamically like liquid 
sulfur\cite{Corrales&Wheeler88,AndersonGreer87,Dup&Pfeuty82}. 

A more direct approach to study self-avoiding polymers was developed by 
Edwards and Des Cloizeaux\cite{Edw65,Clo81,CloJan90}. 
In this approach, hard self-avoidance is replaced by a soft short range 
repulsive interaction between the monomers. 
The repulsive interaction is then studied perturbatively by expanding
about ideal random walks. 
Here too, the perturbative expansion can be reorganized into a renormalization
group about the upper critical dimension 4, which was shown\cite{PGG72} 
to be equivalent to the perturbation expansion of $\varphi^4$-theory in the limit $N\to 0$.  
This equivalence holds both on the formal and on the perturbative levels,
providing  two apparently different approaches for calculating the same exponents. 

There is much work in the field theory community on generalizing results 
for fluctuating lines to entities of other internal dimensions $D$.
The most prominent example is the work on string theories, which describe
$D=2$ world sheets.
An earlier example is provided by the correspondence between gauge theories
and random surfaces\cite{Kogut1979,Savit1980}.
The low temperature expansion of the Ising model in  $d$-dimension also results in a
sum over surfaces that are $d-1$ dimensional.
 For $d=3$, the surfaces are made out of plaquettes, the basic objects
of  lattice gauge theories. 

The simplest generalization of linear polymers is to ``tethered" (or polymerized) 
surfaces\cite{KantorKardarNelson1986a,KantorKardarNelson1986b}, 
which have a fixed internal connectivity, and are 
thus simpler than their gauge theory counterparts. 
 For theoretical analysis, it is convenient to further generalize  to membranes 
of arbitrary (inner) dimension $D$, interpolating between polymers for $D=1$ and 
membranes for $D=2$. 
Simple power counting indicates that the self-avoiding interaction is relevant only for 
dimensions $d<d_c=4D/(2-D)$, making possible an 
$\E=2D-d(2-D)/2\sim (d_c(D)-d)$-expansion, which was
first carried to 1-loop order around this line in Refs.~%
\cite{KarNel87,KarNel88,AroLub87,AronovitzLubensky1988,Duplantier1987}.
To obtain results for polymers or membranes, one now has the freedom
to expand about {\em any} inner dimension $D$, and the corresponding
upper critical dimension of the embedding space\cite{Hwa90}.
This freedom can be used to optimize the calculation of critical exponents.

 Following more rigorous analysis of this novel perturbation series
\cite{Dupal90,DDG1,DDG2,DDG3,DDG4},
recently 2-loop calculations were performed for membranes with 
inner dimension $D$ between 1 and 2
\cite{WieseDavid96b,DavidWiese96a}.
The results have been applied to polymers, where the swelling exponent $\nu$ 
has been found to be 0.59 in an appropriate extrapolation scheme  {\em at both 
1- and 2-loop order}.
By contrast, in  standard field theory, the 1-loop result is an {\em underestimate}, while 
the 2-loop result is an {\em overestimate} by a similar amount. 
For  self-avoiding membranes in 3-dimensional space, 2-loop calculations predict an
isotropic fractal phase with dimension of about 2.4 \cite{WieseDavid96b,DavidWiese96a}.
In addition, there have been  extensive numerical studies 
\cite{PlischkeBoal88,Boaletal1989,AbrEtAl89,HoBaumgaertner1989,HoBaumgaertner1990,%
BaumgaertnerHo1990,Grest91,GrestMurat90,LipowskyGiradet1990,GompperKroll92},
and a few experiments on graphite oxide 
layers \cite{HwaKokufutaTanaka,HWAEXP2,SpectorEtAl94}.

In this article, we reverse the analogy that leads from
the $O(N)$ model to self-avoiding polymers: 
The idea is to generalize the high temperature expansion of the $O(N)$ model
from a gas of self-avoiding loops of fugacity $N$, to a similar gas of closed fluctuating
manifolds of internal dimension $D$.
The primary goal is to obtain a novel analytical
handle on the field theory for $D=1$, and we do not insist that the models for general
$D$ correspond to any physical problem.
Given this caveat, the generalization is not unique.
Encouraged  by its success in polymer theory, we study the generalization to 
{\em tethered} manifolds, and in addition restrict ourselves to the genus of  hyper-spheres.  
 For this class of surfaces calculations are simpler; in particular yielding excellent 
values of the exponent $\nu$ in the limit of polymers\cite{WieseDavid96b,DavidWiese96a}. 
We have chosen hyperspheres as they have no additional anomalous correction 
{\em  exactly at } $D=2$. 
The resulting manifold theory depends on two parameters $N$ and $D$, whose
limiting behaviors reduce to well known models, as indicated in the diagram below.
\begin{figure}[t]\setlength{\unitlength}{1mm}%
\centerline{\begin{picture}(80,50)(0,-10)
\thinlines
\put(0,30){\makebox(0,0){\begin{minipage}{5cm}\begin{center}
${O}(N)$-field theory\end{center}\end{minipage}}}
\put(00,0){\makebox(0,0){\begin{minipage}{5cm}\begin{center}self-avoiding polymers
\end{center}\end{minipage}}}
\put(75,0){\makebox(0,0){\begin{minipage}{5cm}\begin{center}self-avoiding $D$-dimensional tethered membranes\end{center}\end{minipage}}}
\put(75,30){\makebox(0,0){\begin{minipage}{5cm}\begin{center}
${O}(N,D)$-manifold model\end{center}\end{minipage}}}
\put(25,0){\line(1,0){20}}
\put(25,0.04){\makebox(0,0){\,\,\,$<$}}
\put(30,5){\makebox(10,0){$D\to 1$}}
\put(25,30){\line(1,0){20}}
\put(25,30.04){\makebox(0,0){\,\,\,$<$}}
\put(30,35){\makebox(10,0){$D\to 1$}}
\put(0,10){\line(0,1){10}}
\put(0.05,10.8){\makebox(0,0){$\vee$}}
\put(0.05,10.3){\makebox(20,10){$N\to 0$}}
\put(75,10){\line(0,1){10}}
\put(75.05,10.8){\makebox(0,0){$\vee$}}
\put(75.05,10.3){\makebox(20,10){$N\to 0$}}
\end{picture}}
\vspace{3mm}
\caption{Schematic description of the new model, and its limits.}
\label{scheme}
\end{figure}
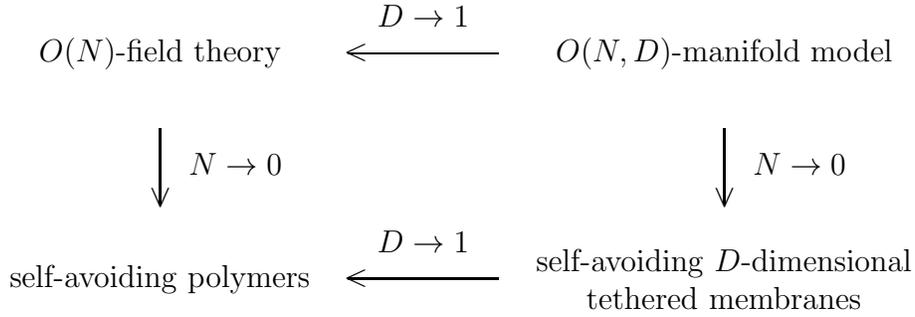

The rest of this paper is organized as follows:
In Sec.~\ref{The O(N)-model in the high-temperature expansion} we review the high-temperature expansion for the $O(N)$ model
to explicitly indicate how it relates the sum over self-avoiding loops.
The different renormalization group schemes used in the literature to
study polymers may potentially lead to some confusion regarding
the  connection to $\varphi^4$-theory.
To clarify the situation, we explicitly compare and contrast the
two main schemes in Sec.~\ref{ren of pol}.
We then generalize the model in Sec.~\ref{Generalization to N colors} to polymers of $N$ different colors, 
making the connection to the $O(N)$ $\varphi^4$-theory. 
Our central result, obtained in Sec.~\ref{Generalization to membranes}, is the first-order expansion for the exponent 
$\nu(D,N,d)$, given by
\begin{equation}\label{central}
\nu(D,N,d) ={2-D\over 2}\times\left[1+{ \E\over 2D}
{1+{c(D)}{N\over2} \over {1\over 2-D} \Gamma\left(\frac D{2-D}\right)^{2}
{\Gamma\left(\frac{2D}{2-D}\right)^{-1}}+1 + c(D) {N\over4}}\right].
\end{equation}
The above result depends on a function $c(D)$, which at the microscopic level 
is related to the relative strengths of self-avoidance between parts of the same manifold, 
and between different manifolds.
In Sec.~\ref{The unknown factor} we propose two choices, $c(D)=1$, and $c(D)=D$, for this parameter.
The numerical values resulting from several extrapolations are discussed in Sec.~\ref{Extrapolations}.
 For the $O(N)$ model in $d=3$, our best extrapolations for the exponent $\nu$
are equal to 0.601, 0.646, 0.676, and 0.697, for $N=0$, 1, 2, and 3 respectively.

The ambiguity associated with $c(D)$ disappears in the $N\to \infty$ limit,
where it is possible to exactly sum the dominant diagrams in the perturbation series.
This result, along with a mean field variational estimate, is presented in Sec.~\ref{The limit N to infty and other approximations}.

The low temperature expansion of the { $d$-dimensional}
Ising model ($N=1$)
provides another route to sums over random surfaces.
As discussed in Sec.~\ref{Low temperature expansions of the Ising model}, the sum is over surfaces of internal dimension 
$D=d-1$, embedded in $d$ dimensions.
However, in crucial difference with tethered manifolds, it is necessary to sum over
all internal metrics (connectivities).
 For $N\to0$, this difference is known to lead to quite drastic geometries.
In particular, sums over a single random surface are dominated by singular
configurations which in fact resemble branched polymers\cite{Cates1988}.
However, for $N\ne 0$, it may be entropically advantageous to break up a 
singular spike into  many bubbles.
If so, a description in terms of fluctuating hyper-spherical surfaces may not be
too off the mark.
The requirement that the dual high and low-temperature expansions of the Ising model
partition function have the same singularity, leads to a putative
identity $\alpha(N=1,d,D)= \alpha(N=1,d,d-D)$.
The numerical tests based upon the 1--loop result of \Eq{central}
appear to support this conjecture.

While obtaining better exponents for the $O(N)$ model is an important goal,
our generalized approach is more valuable if it also applies to other local field theories.
The simplest extension of the $N$-component $\varphi^4$ theory includes cubic anisotropy 
by adding a term proportional to $\sum_\alpha \varphi_\alpha^4$.
The geometrical interpretation of this term is an additional interaction
that only operates between manifolds of the same color.
In Sec.~\ref{Cubic anisotropy} we develop the corresponding manifold extension, whose
renormalization equations involve two interaction parameters, leading to a 
variety of new phases and fixed points, some with Ising criticality. 
A particular scheme in this model yields an Ising exponent of  $\nu^* = 0.6315$ in $d=3$,
which is indistinguishable from the many loop calculations\cite{Zinn}!
The consequences of this generalization for the random bond Ising
model are explored in Sec.~\ref{The random bond Ising model}.
In the standard description with $D=1$ there is no fixed point for the
random bond Ising model at 1-loop order, necessitating  a $\sqrt{\E}$ expansion.
By contrast, we do find a fixed point at this order for $D\neq 1$.

A number of technical discussions are relegated to the appendices:
Appendices A and B present derivations of some properties of the renormalization
group factors used in the text. 
For completeness, and convenience of the reader, some technical
details of dealing with divergences in the perturbation series for $D\neq1$
are presented in Appendices C and D.
Finally, Appendix E deals with the question of what happens if manifolds
of other topology are also considered.

\section{The $O(N)$-model in the high-temperature expansion}
\label{The O(N)-model in the high-temperature expansion}
In this section, we briefly review the high-temperature 
expansion of the $O(N)$ model. (For more extensive reviews, 
see Refs.~\cite{Savit1980} and \cite{DombGreen3}.) The Hamiltonian is 
\be
 {\cal H} = - J N \sum_{\left<i,j \right>} \vec S_i \cdot \vec S_j \ ,
\ee
where the sum runs over all nearest neighbors of a $d$-dimensional
cubic lattice. To obtain the partition function, we have
to integrate over all $\vec S_i$ subject to the constraint
that $|\vec S_i|=1$, resulting in ($K=\beta J$)
\bea \label{partition function}
{\cal Z} &=& \int_{\{\vec S_i\} } \rme^{-\beta {\cal H}}   =\int_{\{\vec S_i\} } \prod_{\left<i,j \right>}
	\rme^{N K\vec S_i \cdot\vec S_j} \ .
\eea
The high-temperature expansion is obtained by expanding 
the exponential factors in \Eq{partition function}  as
\be 
 \rme^{NK\vec S_i \cdot \vec S_j} = 1+ NK\vec S_i \cdot \vec S_j + \cdots \ .
\ee
Typically, only the first two terms in the Taylor-expansion are retained. 
This is justified as we are only interested in universal quantities,
for which the weight is already not unique and 
may be modified %
[$ \exp(NK \vec S_i \cdot \vec S_j)\longrightarrow
 1+NK \vec S_i \cdot \vec S_j $] 
in order to cancel subsequent terms in the Taylor expansion. 

We can represent the various terms in the perturbation expansion in 
the following manner (see Refs.~\cite{Savit1980,KardarLH94}): 
 For each term $NK \vec S_i \cdot \vec S_j$, we draw a line connecting sites $i$ and $j$. 
At any given site $i$, up to $2d$ such lines may join.
The integral over the spin $\vec S_i$ is non-zero, if and only if 
an even number of bonds end at site $i$. For calculational 
convenience, we normalize the integrals by the corresponding solid angle such that  
\be 
	\int \rmd \vec S_i  =1 \ .
\ee
Let us now study the first few terms in the perturbation expansion (see
figure \ref{ht1}).
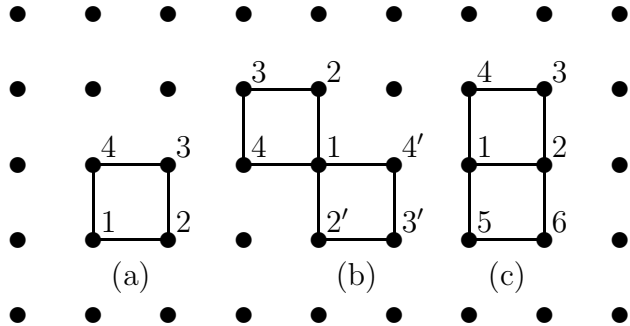
\begin{figure}[htb]\setlength{\unitlength}{1mm}\centerline{\begin{picture}(86,42)(-3,0)
\multiput(0,0)(0,10){5}{\multiput(0,0)(10,0){9}{\circle*2}}
\thicklines
\put(10,10){\multiput(0,0)(0,10)2{\line(1,0){10}}
\multiput(0,0)(10,0)2{\line(0,1){10}}}
\put(30,20){\multiput(0,0)(0,10)2{\line(1,0){10}}
\multiput(0,0)(10,0)2{\line(0,1){10}}}
\put(40,10){\multiput(0,0)(0,10)2{\line(1,0){10}}
\multiput(0,0)(10,0)2{\line(0,1){10}}}
\put(60,20){\multiput(0,0)(0,10)2{\line(1,0){10}}
\multiput(0,0)(10,0)2{\line(0,1){10}}}
\put(60,10){\multiput(0,0)(0,10)2{\line(1,0){10}}
\multiput(0,0)(10,0)2{\line(0,1){10}}}
\put(10,10){\put(1,1){1}\put(11,1){2}\put(1,11){4}\put(11,11){3}}
\put(30,20){\put(1,1){4}\put(11,1){1}\put(1,11){3}\put(11,11){2}}
\put(40,10){\put(1,1){$2'$}\put(11,1){$3'$}\put(1,11){}\put(11,11){$4'$}}
\put(60,20){\put(1,1){1}\put(11,1){2}\put(1,11){4}\put(11,11){3}}
\put(60,10){\put(1,1){5}\put(11,1){6}}
\put(10,0){\makebox(10,10){(a)}}
\put(40,0){\makebox(10,10){(b)}}
\put(60,0){\makebox(10,10){(c)}}
\end{picture}}
\vspace{3mm}
\caption{Some terms in the high-temperature expansion of the $O(N)$-model.}
\label{ht1}
\end{figure}\noindent
The diagram (a) is
\be \label{e5}
	\mbox{(a)}= (KN)^4\int \rmd \vec S_1 \ldots \rmd \vec S_4 \, S_1^\alpha S_2^\alpha  
S_2^\beta S_3^\beta S_3^\gamma S_4^\gamma  S_4^\delta S_1^\delta 
\ .
\ee
To do the integrations, note that \be
	\int \rmd \vec S_i \, \vec S_i^2= \int \rmd \vec S_i \, 1 = 1,
\ee
and therefore
\be
\int \rmd \vec S_i \, S_i^\alpha S_i^\beta = \frac1N \delta^{\alpha\beta}
\ .
\ee
Performing all but the last integration in  \Eq{e5}, we obtain
\be
\mbox{(a)}=K^4 N \int \rmd \vec S_1 \, \vec S_1^2 = K^4 N \ .
\ee
 For any non-intersecting loop, this result is easily generalized 
to
\be
	K^{\mbox{\scriptsize number of links}} N \ ,
\ee 
i.e.\ every closed loop contributes a factor of $N$. Let us now analyze 
what happens when loops intersect and to this aim
 calculate  configuration (b).
Doing all but the integration over $\vec S_1$, we obtain
\be
	\mbox{(b)}=K^8 N^2 \int \rmd \vec S_1 \,  (\vec S_1^2)^2 = K^8 N^2 
=\mbox{(a)}^2\ .
\ee 
Two configurations which have one common point  thus give the same 
contribution in the high-temperature expansion as if they were disjoint.
This is not the case if they have one bond in common, see (c). 
The integral contains an odd power of the field $\vec S_1$, and 
therefore
\be
\mbox{(c)}=0 \ .
\ee

This high-temperature series can  thus be reinterpreted as the sum over all
self-avoiding (non-intersecting) loops. Bonds are {\em totally}\/ self-avoiding, see
e.g.~configuration (c), while vertices are also {\em partially}\/
self-avoiding as can be seen from the following argument. 
There are 3 possible ways to build up configuration (b):
One may take 2 small loops, but there are also 2 possibilities 
to use one loop only. The latter have to be excluded from the partition function.
(There are additional constraints associated with multiple intersections.)
On the other hand, as we are only interested in universal
quantities, taking  precise account  of  these configurations
should be irrelevant as long as bond-self-avoidance is present.
In the direct polymer approach of Edwards and Des Cloizeaux 
\cite{Edw65,Clo81,CloJan90}
discussed below, this corresponds to taking a smaller initial (bare)
coupling constant. 

A single loop can now be viewed as a random walk, i.e.\ as
 the trace of a particle moving 
under Brownian motion. The corresponding Hamiltonian is 
\be \label{bla1}
	{\cal H}_0 = \int_0^L \rmd x\, \frac14 (\nabla r(x))^2  + Lt\ ,
\ee
where $r(x)\in \R^d$ is the trajectory of the particle at time 
$x$ (equivalently, $x$ is the polymer arc-length). 
The total length of the loop is $L=\int \rmd x$.
In addition, one has to demand that the particle returns to
its starting-point, i.e.\ that the polymer is closed. 
To make it  self-avoiding, Edwards and Des Cloizeaux \cite{Edw65,Clo81,CloJan90}
 added
 an explicit 
repulsive interaction upon contact, leading to
\bea 
	{\cal H} &=& \int_0^L \rmd x\, \frac14 (\nabla r(x))^2 
+ \frac {b\mu^\E}4 \int_0^L \rmd  x\int_0^L \rmd y\,  \tilde\delta^d(r(x)-r(y)) +Lt
\label{polymer hamiltonian}
\ .
\eea
The factors of $1/4$, as well as the normalization hidden
in $\tilde \delta$, are chosen for convenience and will be explained later on; 
$\mu$ sets the renormalization scale.
In the high-temperature expansion, there appear loops of all sizes. 
We thus have to sum over all different lengths of the polymer, 
weighted by a chemical potential $t$ conjugate to the length, mimicking
the constant $K$ in \Eq{partition function}. To avoid possible confusion, let
us stress that although closely related, $\ln K$ and $t$ are not identical.
While $K$ is defined as the fugacity for the length of the lattice
walk, the chemical potential $t$ is conjugate to the coarse-grained length.
In principle, the same lattice walk can be represented by curves $r(x)$
of different length $L$.
However, as 
far as universal quantities are concerned, this is unimportant.
Both parameters have to be tuned to reach the critical point, 
and only their deviations from the critical value, but not
the critical value itself, have some physical correspondence.

\section{Renormalization group for polymers}
\label{ren of pol}
We now discuss the perturbation expansion of 
the Hamiltonian in \Eq{polymer hamiltonian}.
Let us start with the correlation functions of the free (non self-avoiding) polymer. 
One has to be careful in distinguishing between open and closed polymers
which will be denoted by subscripts $_o$  and $_c$ respectively.
 For open (or closed, but infinitely long) polymers,  the correlation function
\be
C_o(x)=\frac1d \left< \half ( r(x)-r(0))^2 \right>_o ,
\ee
is the solution of the Laplace equation 
\be \label{open polymer Laplace}
	\half \Delta C_o(x) = \delta(x)
\ ,
\ee
which is easily found to be
\be \label{open polymer propagator}
C_o(x)=|x|
\ .
\ee
 For closed polymers, \Eq{open polymer propagator}
has to be modified. The reason is that the information
has two equivalent ways to travel around a polymer loop of size $L$,
leading to 
\be \label{closed polymer propagator}
C_c(x)= \frac{|x|(L-|x|)}{L}, \qquad \forall\  |x|<L
\ .
\ee

We next calculate the weight of a 
polymer of length $L$. For open polymers this is simply
\be
		\rme^{-Lt}\ ,
\ee
where $t$ is the chemical potential. For closed polymers,
an additional factor of
\be
\left< \tilde \delta^d\left(r(L)-r(0)\right) \right>_{o}
\ee
has to be added, which measures the probability to find
a closed polymer among all open polymers. 
The expectation value therefore is taken with respect to the 
weight for an open polymer, and calculated as follows:
\bea \label{fpp1}
\left< \tilde \delta^d\left(r(L)-r(0)\right) \right>_{o}& =&
\int_k \left< \rme^{ik(r(L)-r(0))} \right>_o \nn\\
&=& \int_k \rme^{- k^2 C_o(L)} \nn\\
&=& \int_k \rme^{- k^2 L} \nn\\
&=& L^{-d/2}
\ .
\eea
The normalizations of $\tilde \delta^d$ and $\int_k$ are chosen for 
calculational convenience such that  
\be
	\int_k \rme^{-k^2 a} =a^{-d/2} \ .
\ee
The same normalizations are also used to incorporate 
self-avoidance as discussed later.
 
To get  the quantities  obtained in the high-temperature
expansion of the loop model introduced above, we still have to 
integrate over all possible lengths of 
the polymer.
We define the partition function for a single polymer as
\be \label{fpp2}
{\cal Z}_1^{(0)} =\int \frac{\rmd L}{L}\, L L^{-d/2} \rme^{-Lt} = 
\Gamma\left(1-\frac d2\right) t^{d/2-1}
\ .
\ee
We have chosen to integrate over a logarithmic scale ($\int \rmd L/L$) in 
order to make the integration measure dimensionless. The factor $L$ counts for
the number of points which may be taken as origin. Our final check, however,
is that we obtain the same result as in the free (Gaussian) field theory.

Additional insight is obtained from a different  way to calculate 
${\cal Z}_1^{(0)}$. If we do not perform the last integral in 
\Eq{fpp1}, \Eq{fpp2} becomes
\bea 
{\cal Z}_1^{(0)} &=&\int \rmd L\, \rme^{-Lt}\int_k\rme^{- k^2 L} \nn\\
 &=& \int_k \frac1{k^2+t} \equiv \,\,\PA
\label{D1}
\ .
\eea
As suggested graphically above, this term of the polymer-perturbation 
theory is equivalent to a term in the  perturbation 
theory of the field-theoretical description of the $O(N)$-model. 
In the usual treatment of the $O(N)$ field theory, the hard constraint of $|\vec S|=1$ is 
replaced in favor of a soft constraint, implemented by the Hamiltonian
\be
{\cal H}_{O(N)} =\int\!\rmd^d r\, \left[ \half (\nabla \vec S(r))^2 + \frac t2  \vec S^2(r) +
\frac{b \mu^\E}{16} (\vec S^2(r))^2 \right] \ .
\ee
 In this description, one has to take the limit
$N\to 0$ in order to allow  for only 1 connected piece. 
(Remember that every closed loop counts a factor of $N$.)
This equivalence, first pointed out  by De Gennes \cite{PGG72},
 is not accidental and can be proven both 
perturbatively  and by formal manipulations of the 
functional integral \cite{Zinn}. 
It reflects the fact that both the field-theoretic formulation of the 
$O(N)$-model, as well as its lattice equivalent, belong to 
the same universality class.
The reader not familiar with this statement is invited to manipulate
a few other terms in the perturbation expansion. 
In the  following discussion, we shall demonstrate this
equivalence for all diagrams encountered. 

We now perform the  perturbation expansion of the polymer Hamiltonian 
in \Eq{polymer hamiltonian}. The first term  is
the expectation value of one $\delta$-interaction 
with respect to the free theory of a closed
polymer, integrated over all positions of the interaction on 
the polymer of length $L$, and then over all polymer-lengths. This is
explicitly 
\bea \label{fpp4}
&&\int_0^{\infty} \rmd L\, L^{-d/2}\rme^{-Lt }
\int_0^{L} \rmd x \int_0^L \rmd y \int_k \left< \rme^{ik(r(x)-r(y))}\right>_c \nn\\
&&\ \ \ =\int_0^\infty \rmd L \,\rme^{-Lt}\int_0^L \rmd x 
\int_0^L \rmd y 
\left[\frac{|x-y| (L-|x-y|)}{ L}\right]^{-d/2} L^{-d/2} \nn\\
&&\ \ \ = 2 \int_0^\infty \rmd L \,\rme^{-Lt}\int_0^L \rmd x 
\int_0^x \rmd z  \left[ z (L-z) \right]^{-d/2} \nn\\
&&\ \ \ = 2\int_0^\infty \rmd z \int_0^\infty \rmd x' \int_0^\infty\rmd y' \, 
\rme^{-t(x'+y'+z)} \int_p\rme^{- p^2 z} \int_k\rme^{- k^2 (x'+y')} \nn \\
&&\ \ \ = 2\int_k \int_p \frac{1}{(k^2+t)^2}\frac1{p^2+t} \nn\\
&&\ \ \ \equiv 2 \ \,\PG = 2\ \, \PH \  \times \ \PA 
\label{e25}
\ .
\eea
The relation to $\varphi^4$-theory is again apparent: 
The integrals in \Eq{fpp4} are ultra-violet divergent. The leading 
 divergence is subtracted via a finite part prescription,
the sub-leading term is treated via  dimensional regularization as a pole in
\be
	\varepsilon = 2 -d/2
\ .
\ee 
(Note the factor of 2 difference from the more usual definition of $\E=4-d$.)

Let us now introduce a renormalized Hamiltonian. Three renormalizations 
may be  required:
A renormalization of the field $r$, of the coupling 
constant $b$, and of the chemical potential $t$. Denoting the bare quantities
with a subscript  $_0$, we set
\bea \label{bla2}
r_0 &=& \sqrt{Z} r  \ ,\nn\\
t_0 &=& Z_t t \ ,\label{Z ren}\\
 b_0 &=&  \mu^{\E}Z^{d/2}Z_b b \nn
\ .
\eea
This yields the renormalized Hamiltonian
\bea 
{\cal H} &=&  Z \int\! \rmd x \frac 1 4 (\nabla r(x))^2 + \frac{b  \mu^\E
Z_b  }4\int\! \rmd x \int\!  \rmd y \,
\tilde \delta^d(r(x)-r(y))
+ Z_t t \int\!\rmd x
\label{H ren}
\ ,
\eea
where $\mu$ sets the renormalization scale. It is possible
to subtract at the scale of the renormalized chemical potential $t$, but 
this turns out to be rather confusing when deriving the renormalization
group equations. 
We can now eliminate the divergence in \Eq{fpp4} by setting 
\bea \label{Zt(N=0)}
Z_t &=& 1- \frac b {2\E} \Res \left( \PA \right)\ .
\eea
This is seen by 
expanding $\rme^{-{\cal H}}$ with ${\cal H}$ given in \Eq{H ren}.
 From Eqs.~\eq{fpp2} and \eq{D1}, we read off the numerical value of $\,\PA\,$, 
yielding
\bea 
 Z_t &=& 1+\frac b {2\E} \ .
\eea

The next step is to study the renormalization of the interaction,
to which the following two diagrams contribute 
\be \label{e30}
\PE\quad,\ \mbox{and}\quad \PF \ .
\ee
To calculate the first diagram, change coordinates to $x_0$ and $y_0$, which indicate
the points midway between the contacts on each polymer. 
The shorter relative distance between these points on each polymer
is denoted by $x$ (or $y$), while the longer one is indicated by
$\Omega_x$ (or $\Omega_y$). The arbitrariness in this choice leads to a 
combinatorial factor of 2 per polymer loop, for an overall coefficient of 4. 
For each contribution of  
\bea
\PE\, &=& 
\int_{k_0,x_0,y_0} \int_{k,x,y} \rme^{-t(\Omega_x+\Omega_y+x+y)}
\left<
\rme^{i(\frac {k_0}2+k)\left(r\left(x_0+\frac x 2\right)-r\left(y_0+\frac y 2\right)\right)}
\rme^{i(\frac {k_0}2-k)\left(r\left(x_0-\frac x 2\right)-r\left(y_0-\frac y 2\right)\right)}
\right>, \nn\\&&
\eea
short distance singularities appear in the integration over $x$ and $y$.
The leading term in the short distance expansion is
\bea
&&
\int_{k_0,x_0,y_0}
\left<\rme^{i {k_0}\left(r\left(x_0\right)-r\left(y_0\right)\right)}\right>
\rme^{-t(\Omega_x+\Omega_y)} 
\int_{k,x,y} \rme^{ -k^2\left(C_c(x)+C_c(y)  \right)}
\rme^{-t(x+y)}.
\eea
 For small arguments, the correlation function can be approximated by
its infinite volume limit, leading up to subleading terms to
\bea
\PD \,\times \int_{k,x,y} \rme^{-(k^2+t)(x+y)} 
 =
\, \PD \,\times \int_k \frac1{(k^2+t)^2}\ .
\eea
The final result is
\bea
&& \,\PE\, =\, \PD \,\times \, \ \PH
\,+\mbox{subleading\, terms}
\ .
\eea
The second diagram in \Eq{e30}
 has already appeared in  \Eq{e25}, and we can symbolically write
\bea
\PF \,= \, \PG  \,\Big/\,  \PA\, =\,\PH = \int_k \frac{1}{(k^2+t)^2}
 = \Gamma\left(2-\frac d2 \right) t^{2-d/2}
\ .
\eea
This diagram appears with a combinatorial factor of 2  for its left-right 
asymmetry, and another factor of 2 for the possibilities to put 
the single point on \,$\PG$\,.

Adding these contributions yields the following renormalization factor at 1-loop order
(note that the combinatorial factors of 4 cancel with that
of the $b/4$ in the Hamiltonian \Eq{H ren}),
\be \label{Zb poly mass}
Z_b= 1+\frac b \E \Res \left(\,\,\PE+\PF\right) = 1 + \frac{2b}\E
\ .
\ee
No field renormalization is necessary ($Z=1$). 

The next step is to calculate the renormalization group functions, 
which measure the dependence of the renormalized quantities upon a
change of  the renormalization scale $\mu$, while keeping the bare values fixed. 
The derivation of these functions is given in appendix \ref{derive RG},
and results in a so-called $\beta$--function
\bea \label{beta pol}
\beta(b)&=& \mu \frac{\partial}{\partial \mu}\lts_0 b =
	\frac{-\E b}{1+b\frac{\partial}{\partial b} \ln Z_b +\frac d2 
b\frac{\partial}{\partial b} \ln Z }
\ ,
\eea
and a scaling function for the field $R$
\be \label{nu pol}
\nu(b)=\half -\half \beta(b) \frac{\partial}{\partial b} \ln \left(Z Z_t\right)
\ .
\ee
We are now in a position to calculate the exponent $\nu^*$ in 1-loop order.
The $\beta$-function is at this order 
\bea
	\beta(b) &=& -\E b +b^2 \Res \left(\,\, \PE+ \PF\right)
 +{\cal O} (b^3) + {\cal O}(b^2\E)\nn\\
	&=& -\E b +2 b^2 +{\cal O} (b^3) + {\cal O}(b^2\E) \  ,
\eea
and the scaling function $\nu(b^*)$ becomes
\bea
\nu(b^*) &=& \half -\half \Res \left(  \PA  \right) \frac{b^*}2 +{\cal O}(\E^2)\nn\\
&=& \half -\frac\E4 \frac {\Res\left( \PA\right) } {\Res \left( \,\,\PE+\PF\right)} 
	+{\cal O} (\E^2) \nn\\
&=& \half +\frac \E 8 + {\cal O}(\E^2) 
\ .
\eea

This renormalization scheme is also used in $\varphi^4$-theory. At 1-loop order, no renormalization 
of the wave-function is necessary. Only the reduced
``temperature'' $t$ is renormalized.
There is another scheme, equally useful, to perform the 
renormalization of polymers, which is also used in the broader context
of polymerized membranes. This scheme also works  for infinite
membranes. Naturally, for infinite membranes, no renormalization 
of $t$ can occur as it is  identically 0.
It is also  known 
for the renormalization of standard field-theories that one has 
 the choice to work either in a massive ($t\ne0$) or a massless ($t=0$)
scheme. 

For the polymer model, let us find a renormalization scheme where $t$ is not 
renormalized, and therefore the limit $t\to 0$ can be taken without problem.
The key observation is that only the combinations $ZZ_t$ and
$Z_bZ^{d/2}$ enter the renormalization group calculations, and these 
combinations are left invariant by changing the $Z$-factors to
\bea
Z_t' &=& 1 \ ,\nn\\
\label{Rin1}
Z'&=& Z Z_t \ ,\\
Z_b'&=&Z_b Z_t^{\E -2} \ .\nn 
\eea
For a derivation of this property as a consequence of the 
rescaling-invariance of the underlying Hamiltonian, see Appendix \ref{rescale}.
In terms of the modified $Z$-factors, we obtain
\bea
\beta(b)&=&\frac{-\E b}{1+b \frac{\partial}{\partial b} \ln Z_b' + \frac d2
b\frac{\partial}{\partial b} \ln Z'} \ ,\nn\\
\nu(b) &=& \half - \half \beta'(b) \frac{\partial}{\partial b}
\ln Z'
\ .
\eea
This is the scheme used by David, Duplantier and  Guitter \cite{DDG3,DDG4},
 and by David and Wiese \cite{WieseDavid96b,DavidWiese96a} in the 
context of polymerized membranes, where it is  
the most suitable for higher loop calculations.
On the other hand, it may lead to some confusion as it necessitates a 
renormalization of the field, even in the case of polymers. 
This may not have been expected from the 1 to 1 correspondence on 
the level of diagrams for the 
$N\to0$ limit of $\varphi^4$-theory, and polymers. 
As shown above, the two schemes are completely 
equivalent and one may use the one better suited to the problem at hand.

Let us stress another important difference 
between the two approaches. This is most easily done by 
using the multilocal operator product expansion (MOPE) introduced
in Refs.~\cite{DDG3,DDG4}, and heavily used in Refs.~\cite{WieseDavid95,WieseDavid96b,DavidWiese96a}. 
To this end, let us write the multilocal operator
$\tilde \delta^d(r(x)-r(y))$ as $_x\GB_y$. Divergences in the perturbation 
expansion then occur when distances become small. The first such 
configuration  is the contraction of the end-points of one dipole, which 
we shall denote by $\GH$. 
A derivation of the contribution of this diagram in the more general context 
of self-avoiding membranes is given in appendix \ref{Div and MOPE}.  
Specializing to polymers gives the result
\be
_x\GH_y = |x-y|^{-d/2} \mbox{\bf 1} -\half |x-y|^{1-d/2}  \GO + \ldots\ ,
\ee
where
\be
	\GO := \half (\nabla r)^2
\ .
\ee
The divergence proportional to the operator \mbox{\bf 1} is subtracted
by analytical continuation. In the absence of any boundary and for 
infinite membranes,  this term has no effect on the renormalization 
functions. The second term is more serious and has to be subtracted. This
is done by renormalization of the field, thus introducing the renormalization factor
\be
 Z= 1+ \frac b {2\E}
\ .
\ee
Upon expanding the Hamiltonian, this  yields a counter-term 
proportional to $\GO$, which cancels the divergence. 

Let us now study the renormalization of the coupling constant
in this scheme. Using the MOPE, we can write down the following 
two UV-divergent configurations
\be
	\GM \ ,\qquad \mbox{and} \qquad \FD \ ,
\ee
from which we shall extract terms proportional to the interaction 
$\GB$, which we denote as 
\be
 \bigg< \GM \bigg | \GB \bigg>\ ,\quad  \mbox{and} \quad \bigg<\FD\bigg | \GB \bigg>
\ .
\ee
The first is written in the notation of polymer theory as 
\be \label{first div b}
	\bigg< \GM \bigg | \GB \bigg> = \PE \ .
\ee
Indeed  this diagram was subtracted when we renormalized the interaction in
 \Eq{Zb poly mass}, where we also subtracted  the term, 
\be \label{second div b}
\quad \bigg<\FD\bigg | \GB \bigg> = \PF \ \ .
\ee
The MOPE now tells us that 
\be \label{second div b '}
\FD = \GB \times \GH \ ,
\ee
as will also be proved in the context of membranes
(see appendix \ref{Div and MOPE}, or 
Refs.~\cite{DDG3,DDG4,WieseDavid96b}, where this
is discussed in some detail). 
This result implies that having introduced a 
counter-term for $\GH$, i.e.\ a renormalization of the field, no
 counter-term for the diagrams in \Eq{second div b} is needed.

We can check for consistency by comparing
the $\beta$-functions from the two schemes at 1-loop order.
In the massive scheme, we had 
\bea \label{beta1}
\beta(b)&=& -\E b + b^2 \, \Res \left(\,\, \PE + \PF \right) 
+	{\cal O} (b^3) +  {\cal O} (b^2\E) \ .
\eea
In the massless scheme, we obtain
\bea \label{beta2}
\beta(b)&=& -\E b + b^2 \, \left( 
\Res  \bigg< \GM \bigg | \GB \bigg> 
- \frac d 2 \Res 
\bigg< \GH \bigg | \GO \bigg> \right)+
{\cal O} (b^3) +  {\cal O} (b^2\E) 
\ .\qquad
\eea
It is now easy to see that expressions \eq{beta1} and \eq{beta2} are equivalent up to order ${\cal O} (b^3)$ and  ${\cal O} (b^2\E)$,
since
\bea
&&\Res\bigg< \GM \bigg | \GB \bigg>= \Res \left(\,\,\PE\,\,\right)
=\Res\left(\,\,\PF\,\right) =1 \ ,
\eea
and
\be
-\frac d2 \Res \bigg< \GH \bigg | \GO \bigg> =\frac d 4 =1 +{\cal O}(\E)
\ .
\ee
Another observation is that in the massless scheme, vertex
operators like $\rme^{ik(r(x)-r(y))}$ are finite, whereas in 
the massive scheme they require an additional renormalization.

\section{Generalization to $N$ colors}
\label{Generalization to N colors}

Having performed a careful analysis of the different
renormalization schemes, we are now in a position to
generalize to the  case $N>0$, i.e.\ to  an 
arbitrary number of self-avoiding polymer loops.
To this aim, we introduce polymers of $N$ different colors, and for the
time-being,  work in the massive scheme.
In addition to $\,\,\PG\,\,$, which renormalizes the 
chemical potential $t$, there is now a second contribution,
namely  
\be
\PB\ \ .
\ee
This diagram is easily factorized as
\be
\PB \ = \ \PH \ \times \ \PA \ ,
\ee
and is therefore equivalent to the digram already encountered in \Eq{fpp4}
and absorbed in $Z_t$ (for $N=0$ in \Eq{Zt(N=0)}).

Let us now determine the combinatorial factor: A configuration 
\be
	\PI
\ee
can be made out of 1 polymer in 2 different ways or out of two polymers.
The latter comes with an additional factor of $N$, accounting for the $N$ 
different colors introduced above. 
$Z_t$ is thus modified to
\bea \label{Zt mod}
Z_t &=& 1- \frac{b}{2\E} \Res\left( \PA \right) \left( 1+\frac N2 \right) 
+{\cal O} (b^2) \nn\\
&=& 1+\frac b{2\E} \left( 1+\frac N2 \right)+{\cal O} (b^2) 
\ .
\eea
This is indeed the same combinatorial factor as derived 
from $N$-component $\varphi^4$-theory. 
For the renormalization of the coupling-constant, in addition to 
\be \label{prop 1}
	\PE \ ,\qquad \mbox{and} \qquad \PF \ ,
\ee
there is  the possibility that an additional loop 
mediates the interaction between two given polymers, described by a configuration
\be \label{prop N}
	\PC
\ .
\ee
The configurations in \Eq{prop 1} are realized in 4 different ways each in the 
high-temperature expansion, while for \Eq{prop N} there is only one realization
which comes with a  factor of $N$ for the $N$ different colors. 
$Z_b$ is therefore modified to 
\bea 
Z_b&=& 1+\frac b \E \Res \left(\,\,\PE +\PF 
 +\frac N4\,\,\PC\,\right) \nn\\
&=& 1 + \frac{b (8+N)}{4\E} 
\label{Zb poly mass N}
\ .
\eea
Evaluating the critical exponent $\nu^*$ as before now yields
\be
\nu^* = \half +\frac\E2 \frac{2+N}{8+N}
\ .
\ee
It is again possible to switch to the massless scheme.  At this stage 
this is not very enlightening, as for polymers all diagrams are essentially
equivalent. We will therefore discuss this scheme in the context of membranes,
which is introduced in the next section.

\section{Generalization to membranes}
\label{Generalization to membranes}
We shall now introduce a generalization to polymerized tethered membranes.
 Formally, one
 generalizes the function $r(x)$
to \cite{KantorKardarNelson1986a,KantorKardarNelson1986b,KarNel87,KarNel88,AroLub87,AronovitzLubensky1988,Dupal90,DDG1,DDG2,DDG3,DDG4,WieseDavid96b,DavidWiese96a,WieseDavid95}
\be
	r: x\in \R^D \to r(x) \in \R^d \ ,
\ee%
{\begin{figure} 
\epsfxsize=0.7\textwidth \centerline{\epsfbox{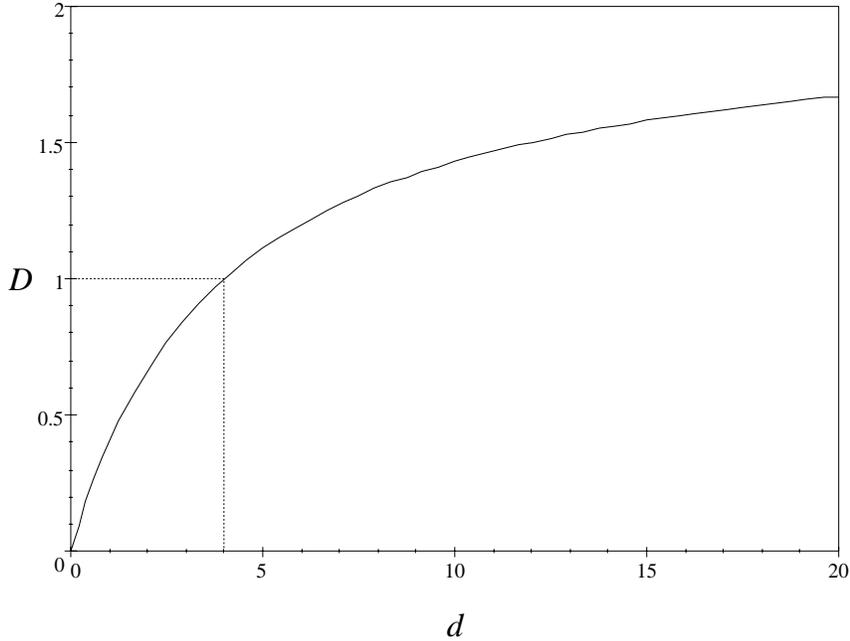}}
\caption{The critical curve $\E(D,d)=0$. The dashed line corresponds to the 
standard polymer perturbation theory, critical in $d=4$.}%
\label{f:kritdim}%
\end{figure}}%
with membranes obtained from $D=2$. While perturbation 
theory is always singular for $D=2$, it is  possible to 
perform an analytical continuation in the inner dimension for $0<D<2$. 
It is now possible to make an $\E$-expansion, where
\be
\E=2D -\nu_0 d \ ,\qquad \mbox{with}\qquad \nu_0=\frac{2-D}2 \ ,
\ee
about {\em any point} $(D,d)$ for which $\E=0$ (see Fig.~\ref{f:kritdim}). 
(The final results are then evaluated for $D=2$ or 1.) 

We shall re-write the Hamiltonian as
\bea \label{Hamiltonian}
{\cal H} &=& \frac Z{2-D} \int_x \half (\nabla r(x))^2 
+ b \mu^{\E}
 Z_b   
\int_x\int_y \tilde \delta^d(r(x)-r(y)) + t Z_t \Omega \ ,
\eea
where the normalization of the integration measure has been chosen for 
convenience such that
\be
 \int_x := \frac1{S_D} \int \rmd^D x\ , 
\qquad \mbox{where}\qquad S_D=\frac{2\pi^{D/2}}{\Gamma(D/2)}
\ .
\ee
 For radial symmetric functions, the integration is then simply
\be
\int_x f(|x|) =\int_0^\infty \frac {\rmd x}x \, x^D \, f(x) \ ,
\ee
and the volume of the membrane is
\be
\Omega = \int \rmd^Dx = S_D \int_x \ .
\ee
Note that for $D=1$, normalizations agree with those used for polymers
in the preceding sections. 

The factor $1/(2-D)$ is introduced in \Eq{Hamiltonian} in order to obtain
\be \label{free cor}
	C_0(x-y) := \frac1d \left< \half\left(r(x)-r(y)\right)^2 \right> = 
|x-y|^{2-D} +\,\mbox{subleading terms} \ .
\ee
To calculate the next to leading term in the free correlator of \Eq{free cor},
note that for any function $C_0(x)$ defined on a closed compact manifold 
\be
\int_x \Delta C_0(x) = 0 \ .
\ee
 For the correlator to satisfy the above condition, the usual Laplace equation, 
$\Delta C_0(x) \sim \delta^D(x)$,  has to be modified to
\be
 \frac{1}{(2-D)S_D} \Delta C_0(x) = \delta^D(x) -\frac1\Omega \ ,
\ee
where $\Omega$ is the volume of the compact manifold.
The numerical prefactors come from our choice of normalizations
in \Eq{Hamiltonian}. In the infinite-volume limit, the correction term disappears, 
and the usual equation is regained. 
It is easy to deduce that
\bea \label{free cor+2}
C_0(x)=	\frac1d \left< \half(r(x)-r(y))^2 \right> &=& |x-y|^{2-D} -
\frac{\nu_0 S_D}{D\Omega} |x-y|^2 
+ \mbox{subleading terms} \ .\qquad 
\eea
The coefficient of the correction term clearly agrees for $D=1$
with the exact result for closed polymers in \Eq{closed polymer propagator}.

The considerations of section \ref{ren of pol} can now be
generalized to the case of membranes. 
The free partition function for a single polymer, i.e. the sum over all sizes of 
a non-interacting polymer,  \Eq{fpp2},
is  generalized to 
\bea 
{\cal Z}_1^{(0)} &=& \frac{c(D)}{D} \int \frac{\rmd \Omega} \Omega\, \Omega\, \Omega^{-\nu_0 d/D} \rme^{-t \Omega} \nn\\
&=& \frac{c(D)}D \Gamma\left( \frac\E D -1 \right) t^{\E/D-1}
\label{Z1(0)mem} 
\ .
\eea
We have chosen to integrate over a logarithmic scale, 
$\frac{\mbox{\scr d} x}x = \frac{1}{D} \frac{\mbox{\scr d} \Omega}{ \Omega} $.
To emphasize the arbitrariness of this choice, we have included an additional 
factor of $c(D)$, which is further discussed in the next section.
This factor is important, as it also appears in the ratio of 
divergences due to self--interactions of one membrane, and those 
of interactions with other membranes. 
The factor $\Omega$ in the integrand of the above equation 
originates from the possible choices of 
a point $x_0$ on the membrane, while the factor 
\be 
\Omega^{-{\nu_0 d/D }} \sim \left< \tilde \delta^{d}(r(x_0)) \right>_0 \ ,
\ee
is the probability that at this point the membrane is attached to a given point in space.
As usual, we have introduced a chemical potential proportional to the size of 
the membrane.

Let us now generalize \Eq{e25}  for the effect of one $\tilde \delta^d$-insertion 
from the expansion of the interaction. For the time being, we fix the size of the 
membrane to $\Omega$, and evaluate
\be
\int_x \int_y \left< \tilde \delta^d(r(x)-r(y)) \right>_0 \ . \ee
This integral is (see \Eq{fpp1})
\be
\frac{\Omega}{S_D} \int_x  C_0(x)^{-d/2} \ ,
\ee
and we have to remove all UV-divergent contributions. To do so, we expand 
$C_0(x)^{-d/2}$ for small $x$. Up to UV-convergent terms, this is 
(using \Eq{free cor+2})
\be
\frac{\Omega}{S_D} \int \frac{\rmd x}x x^D \left(x^{-\nu_0 d}+
\frac d2  \frac{\nu_0 S_D}{D\Omega} x^{D-\nu_0 d} +\ldots \right)
\ .
\ee 
The first term is strongly  UV-divergent and has to be subtracted by a finite part 
prescription, while the second is (up to terms of order $\E^0$) equal to
\be \label{mem1}
 \frac1\E \Omega^{\E/D}
\ .
\ee
Note that we have cut off the integral at the upper bound $x_{\mbox{\scriptsize max}}=
\Omega^{1/D}$. This procedure may appear rather crude, but the residue of the pole 
in $1/\E$ is not affected \cite{WieseDavid95}. 

Upon integrating over all scales, the partition function (to first order) reads
\bea 
{\cal Z}_1^{(1)} &=& \frac{c(D)}{D} \int \frac{\rmd \Omega} \Omega\, \Omega\, \Omega^{-\nu_0 d/D} \rme^{-t \Omega} \left[  1+\frac b\E (\Omega t)^{\E/D} + \ldots \right]
\label{Z1(1)mem} 
\ ,
\eea
which upon integration over $\Omega$ results in
\be \label{X}
{\cal Z}_1^{(1)} = {\cal Z}_1^{(0)} \left[  1+\frac b{2\E}  + \ldots \right]
\ee
Note the difference in factor of 2 between Eqs.~\eq{mem1} and \eq{X},
which is due to a subtlety known to result from  nested integrations
in standard
 field theories. 
This factor of 2 
can also be interpreted as being geometric. The counter-term is only 
needed in the half-sector $x<\Omega^{1/D}$ and not in the 
half-sector $x>\Omega^{1/D}$.
(See also the calculations in Refs.~\cite{Dupal90,WieseDavid96b}).
Introducing now a counter-term for $t$  yields
\be \label{mem2}
Z_t= 1 + \frac b{2\E}  \ .
\ee
The bare and renormalized quantities are now related by
generalizing \Eq{bla2} to
\bea 
r_0 &=& \sqrt{Z} r\ , \nn\\
t_0 &=& Z_t t \ ,\\
b_0 &=& \mu^\E Z^{d/2}Z_b b \nn
\ ,
\eea
leading to the renormalization group functions
(compare with Eqs.~\eq{beta pol} and \eq{nu pol})
\bea \label{RG-functions}
\beta(b)&=&\mu\frac{\partial }{\partial \mu}\lts_0 b	= \frac{-\E b}{1+b\frac{\partial}{\partial b} \ln Z_b +\frac d2 
b\frac{\partial}{\partial b} \ln Z }
\ ,\\\label{RG-functions'}
\nu(b)&=&\frac{2-D}{2} -\half \beta(b) \frac{\partial}{\partial b} \ln \left(Z Z_t^{(2-D)/D}\right)
\ .
\eea
(The derivation is given in appendix \ref{derive RG}.)

The combinations $ZZ_t^{(2-D)/D}$ and $Z_bZ^{d/2}$,
which  enter the renormalization group calculations,  are left invariant 
by changing the $Z$-factors to%
\bea
Z_t' &=& Z_t/Z_\alpha \ ,\nn\\
Z'&=& Z Z_\alpha^{(2-D)/D} \ ,\\
Z_b'&=&Z_b Z_\alpha^{\E/D-2} \ .\nn
\eea
For a derivation of this property as a consequence of the 
rescaling-invariance of the underlying Hamiltonian, see Appendix \ref{rescale}.

In order to eliminate the renormalization of $t$, we chose
\be
Z_\alpha=  Z_t \ ,
\ee
resulting in 
\bea
Z_t' &=& 1 \ ,\nn\\
\label{Rin2}
Z'&=& Z Z_t^{(2-D)/D} \ ,\\
Z_b'&=&Z_b Z_t^{\E/D-2} \ ,\nn
\eea
and the renormalization group functions
\bea
\beta(b)&=&\frac{-\E b}{1+b \frac{\partial}{\partial b} \ln Z_b' + \frac d2
b\frac{\partial}{\partial b} \ln Z'} \ ,\nn\\
\nu(b) &=& \frac{2-D}2 - \half \beta(b) \frac{\partial}{\partial b}
\ln Z'
\ .
\eea
With this change of variables,  \Eq{mem2} is replaced by $Z_t'=1$, and
\be
Z '= 1+ \frac{2-D}{2D} \frac b\E
\ .
\ee

The above result is precisely that obtained by using the multilocal operator 
product expansion technique for infinite membranes (see appendix \ref{Ren for
inf mem} 
and Ref.~\cite{WieseDavid96b}),
where the renormalization factor is calculated from
\be
	Z'=1-(2-D) \frac b \E \Res \bigg< \GH \bigg | \GO \bigg>, \qquad\mbox{with}
\qquad \Res \bigg< \GH \bigg | \GO \bigg> = -\frac1{2D}
\ .
\ee
The interpretation of this formula is simple, as
\be
\bigg< \GH \bigg | \GO \bigg>
\ee
is just the diverging contribution form the MOPE, of one $\tilde \delta^d$-insertion. 
 For $N=0$, the renormalization of the coupling constant in the 
massless scheme is analogously 
(see section \ref{ren of pol}, appendix \ref{Ren for inf mem}, and
Ref.~\cite{WieseDavid96b})
\bea
&&Z_b'=1 + \frac b\E \Res\bigg< \GM \bigg | \GB \bigg> \ ,
\eea
with
\bea
&&\Res\bigg< \GM \bigg | \GB \bigg> =\frac1{2-D} \frac{\Gamma\left(\frac D{2-D}\right)^{\!\!2}}
{\Gamma\left(\frac{2D}{2-D}\right)}
\ .
\eea
Alternatively, in the massive scheme ($Z=1$),
\bea
Z_b &=& \left( 1 + \frac b\E \Res\bigg< \GM \bigg | \GB \bigg>\right)
 \times Z_t^2 \nn\\
	&=& 1 + \frac b\E \left(\Res\bigg< \GM \bigg | \GB \bigg>  +1\right)
\ .
\eea

Let us now study the generalization to $N$ components  in  
the massive scheme. 
Taking care of the additional factor of $c(D)$ introduced in the definition of 
the free partition function  in \Eq{Z1(0)mem},   \Eq{mem2}
is modified in the same manner as \Eq{Zt mod} to 
\be
Z_t= 1 + \frac b{2\E} \left(1+ \frac{c(D)N}2 \right) 
\ .
\ee
There are several possibilities to derive the modification to 
the renormalization factor of the interaction. For a direct 
derivation  generalizing \Eq{Zb poly mass N} we
 calculate the diagram $\PH$ for membranes, as
\be
\PH = \frac{c(D)}{D} \int \frac{\rmd \Omega}{\Omega} \, \Omega^2 \,
\Omega^{-\nu_0 d} \rme^{-t \Omega}
\ .
\ee
We already have given the derivation of   a similar integral in
\Eq{Z1(0)mem}. The only difference is that now a second
 factor $\Omega$ appears
to take into account the additional point which moves on the membrane.
Integration over $\Omega$ yields
\be
\PH = \frac{c(D)}{D} \Gamma\left(\frac\E D\right) t^{-\E/D} = c(D) \frac1\E t^{\E/D} +{\cal O }(\E^0) 
\ .
\ee
This term again appears with a relative combinatorial factor of $N/4$ compared to the 
other contributions, as discussed in the polymer case. 
The renormalization factor $Z_b$ therefore becomes
\be
Z_b= 1 + \frac b\E \left(\Res\bigg< \GM \bigg | \GB \bigg>  +1 + \frac{c(D) N}4\right)
\ .
\ee 
It is now easy to derive the renormalization group functions
\bea
\beta(b)&=& -\E b + b^2 \left(\Res\bigg< \GM \bigg | \GB \bigg>  +1 + \frac{c(D) N}4\right) 
+{\cal O} (b^3) + {\cal O} (b^2\E)  \ ,\qquad
\eea
and
\bea
\nu(b) &=&\frac{2-D}2 \left(1+\frac b{2D}  \left(1+\frac{c(D)N}2 \right) \right) +{\cal O} (b^2)
\ .
\eea
At the non-trivial (IR-stable) fixed point, this yields the critical exponent
to order $\E$
\bea \label{final result}
\nu^*&=&\frac{2-D}2 
\left(1+\frac \E{2 D} \frac{ 1+{c(D)N}/2  }
{\Res\bigg< \GM \bigg | \GB \bigg>  +1 + c(D) N/4} \right) 
\ ,
\eea
representing our central result for  the generalized $O(N)$-model,
also discussed in the introduction in \Eq{central}.

\section{The arbitrary factor $\lowercase{c} (D)$}
\label{The unknown factor}
In calculating the free partition function in \Eq{Z1(0)mem},
we introduced an arbitrary factor of $c(D)$. 
In principle, any function of $D$ which satisfies
\be
c(1)=1 \ ,
\ee
reproduces the correct result for linear objects.
The additional freedom (or ambiguity) is apparently a reflection
of the non-uniqueness of the generalization to manifolds.
Even after restricting to the class of hyperspheres, there is a
remaining ambiguity in the choice of the measure for the size
of these manifolds.
This arbitrariness carries over to our generalization of the $O(N)$ 
model to $N$-colored membranes. 
(Note also that $c(D)$ is independent of the introduction of 
factors like the $\frac1{2-D}$ in \Eq{Hamiltonian}.)

We shall focus on two natural choices for $c(D)$. 
The first possibility  is to demand that 
the free partition function in \Eq{Z1(0)mem},
\be
{\cal Z}_1^{(0)} = 
\frac{c(D)}D \Gamma\left( \frac\E D -1 \right) t^{\E/D-1},
\ee
has the simplest possible form in the sense of depending on only
one parameter besides $t$, namely $\E/D$. 
This implies 
\be \label{c(D)=D}
c(D)=D \ ,
\ee
which is our favorite choice, and equivalent to the measure $\rmd \Omega/\Omega$
over all scales. 

Another reasonable choice is to demand that the ratio of the 
 probability that one membrane touches itself, to the probability
 that two membranes 
touch, which is 
\be
4/Nc(D), 
\ee
be independent of
$D$. This leads to 
\be \label{c(D)=1}
	c(D)=1\ ,
\ee
and is equivalent to the measure $\rmd x/ x $ over scales.
We shall study both choices in the next section.

\section{Extrapolations}
\label{Extrapolations}
{\tabcolsep1.4mm\begin{figure}[b]\centerline{\renewcommand{\arraystretch}{1.20}
\begin{tabular}[t]{|l|c|c|c|c|c|} \hline
$\mbox{Expanded quantity}$ & $c(D)$ &  $N$& $\nu$
\\ \hline \hline 
$\nu d$ &--- & 0 & 0.76 \\ \hline
$\nu d$ & $1$ & 1 & 0.87 \\ \hline
$\nu d$ & $D$ & 1 & 0.91 \\ \hline
$\nu d$ (lin. in $N$) &$D$ &  1 & 0.96 \\ \hline \hline
 exact & --- &   0 & 0.75  \\ \hline
exact  & --- & 1 & 1  \\ \hline
\end{tabular}\renewcommand{\arraystretch}{1.0}}\vspace{2mm}
\caption{Results of the extrapolations for $\nu$ for polymers and
the Ising model in two dimensions.}%
\label{nutab1}%
\end{figure}}
{\tabcolsep1.4mm\begin{figure}[tb]\centerline{\renewcommand{\arraystretch}{1.20}%
\begin{tabular}[t]{|c|c|c|c|c|} \hline
$\mbox{Expanded quantity}$ & $c(D)$ &  $N$& $\nu$, our result & $\nu$, from \cite{Zinn}
\\ \hline \hline 
$\nu d$ & $D$ &  0 & 0.601 & 0.589 \\ \hline
$\nu d$ & $D$ &  1 & 0.646 & 0.631 \\ \hline
$\nu d$ & $D$ &  2 & 0.676 & 0.676 \\ \hline
$\nu d$ & $D$ &  3 & 0.697 & 0.713 \\ \hline
\end{tabular}\renewcommand{\arraystretch}{1.0}}\vspace{2mm}
\caption{Results of the extrapolations for $\nu$ for the 
$O(N)$ model in 3 dimensions.}%
\label{nutab2}%
\end{figure}}
{\tabcolsep1.4mm\begin{figure}[tb]\centerline{\renewcommand{\arraystretch}{1.20}%
\begin{tabular}[t]{|l|c|c|c|c|} \hline
$\mbox{Expanded quantity}$ & $c(D)$ & $N$& $\nu$, our results & $\nu$, from \cite{Zinn}
\\ \hline \hline 
$\nu (d+2) $  & $D$ & 0 & 0.593 & 0.589 \\ \hline
$\nu (d+2)$ (lin.\ in $N$)& $D$ & 1 & 0.637 & 0.631 \\ \hline
$\nu (d+2)$ (lin.\ in $N$)& $D$ & 2 & 0.681 & 0.676 \\ \hline
$\nu (d+2)$ (lin.\ in $N$)& $D$ & 3 & 0.721 & 0.713  \\ \hline
\end{tabular}\renewcommand{\arraystretch}{1.0}}\vspace{2mm}
\caption{Results of the extrapolations for $\nu$ for the 
$O(N)$ model in 3 dimensions, linearized in $N$.}%
\label{nutab3}%
\end{figure}}
\begin{figure}[htb]\centerline{
\epsfxsize=0.7\textwidth \parbox{0.7\textwidth}{\epsfbox{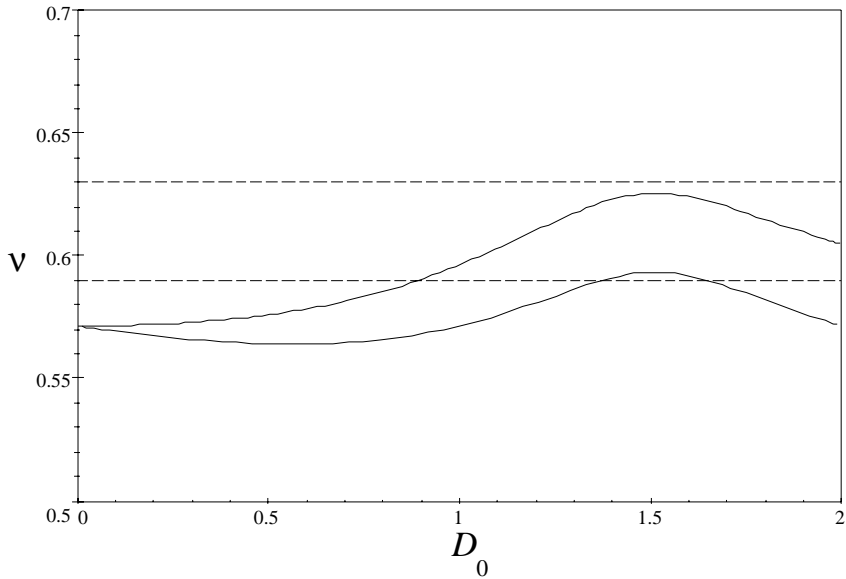}}
}
\caption{Extrapolations for $\nu$ from the expansion of $\nu(d+2)$ to 
$(D=1,d=3)$, for $N=0$ (lower curve) and $N=1$ (upper curve). The straight
lines are a guide for the eye and represent the known results of 0.59 and
0.63 respectively.}
\label{extr1}
\end{figure}\begin{figure}[htb]\centerline{
\epsfxsize=0.7\textwidth \parbox{0.7\textwidth}{\epsfbox{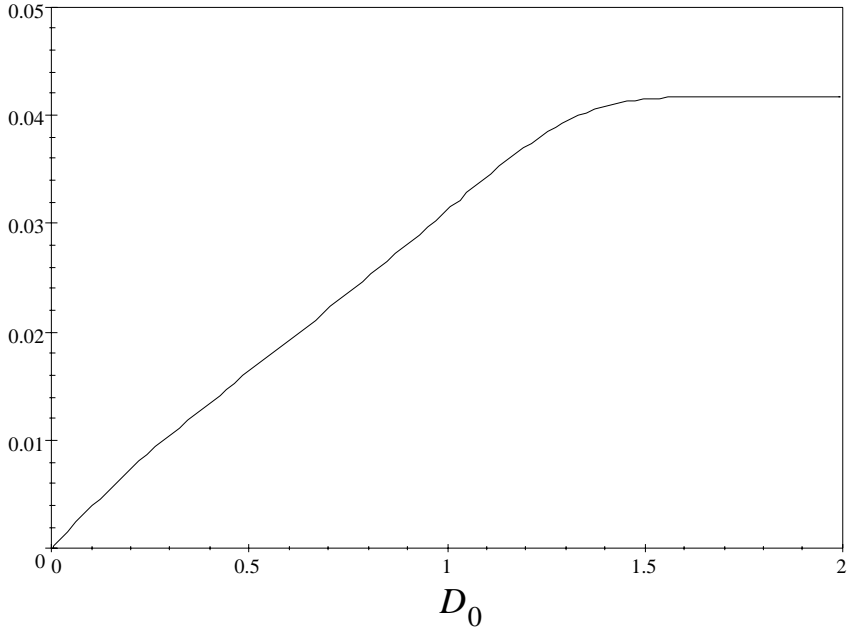}}
}
\caption{Extrapolations for $\frac{\delta}{\delta N}\nu(N)$ from the expansion of $\nu d$ to 
$D=1$, $d=3$. This yields the flattest plateau encountered in all
extrapolations, and there is no difference to calculations in higher loop
order (see \protect\cite{Zinn}).}
\label{extr2}
\end{figure}
\begin{figure}[htb]
\centerline{
\epsfxsize=0.7\textwidth \parbox{0.7\textwidth}{\epsfbox{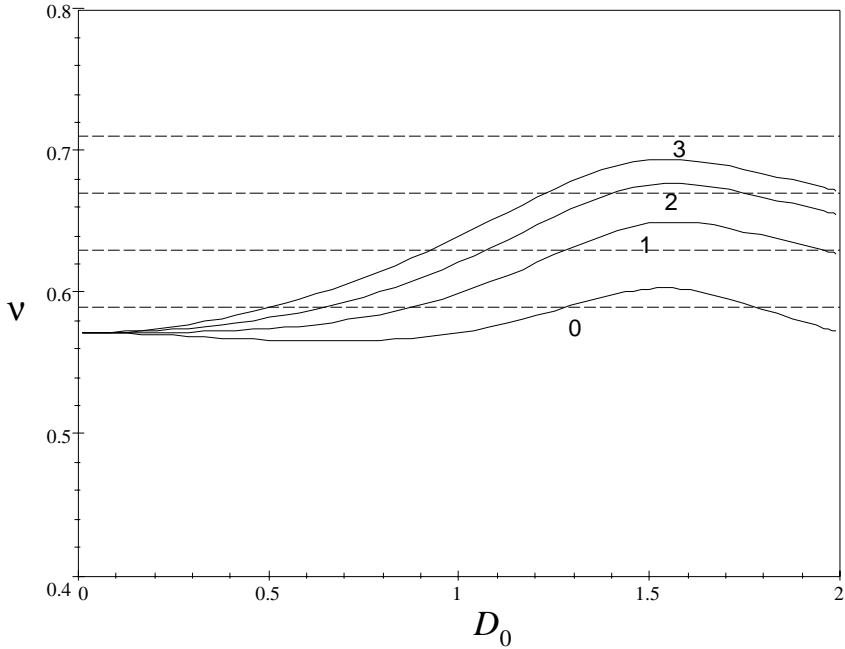}}
}\caption{Extrapolations for $\nu$ from the expansion of $\nu d$ to 
$D=1$, $d=3$, $N=0,\ldots,3$, $c(D)=D$.
The dashed lines are marks for the eye and represent the exact values.}%
\label{4 curves}%
\end{figure}
\begin{figure}[htb]
\centerline{
\epsfxsize=0.6\textwidth \parbox{0.6\textwidth}{\epsfbox{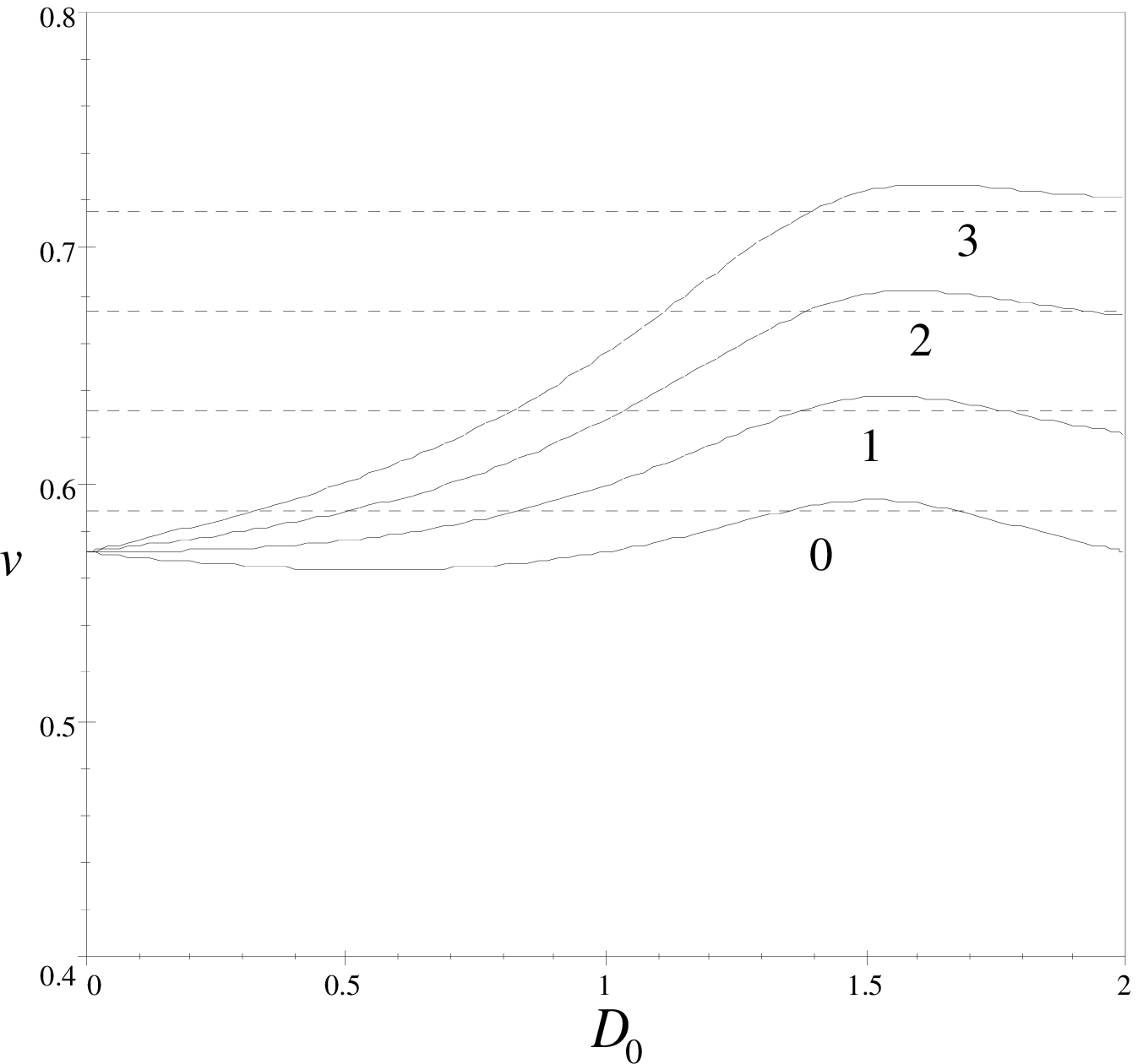}}
}\caption{Extrapolations for $\nu$ after linearization in $N$ 
from the expansion of $\nu (d+2)$ to 
$(D=1,d=3)$ for $N=0,\ldots,3$, with $c(D)=D$.
The dashed lines are marks for the eye and represent the exact values.}%
\label{4 curves'}%
\end{figure}
This section is devoted to extracting the information about the exponent
$\nu$ in physical dimensions, from the general result in  \Eq{final result}.
(We shall use $\nu$, rather than $\nu^*$, to denote the fixed point value.)
As will become apparent, various extrapolation schemes are possible,
and choosing the best one is almost an art; we shall rely heavily on the 
methods  developed in Ref.~\cite{WieseDavid96b} to which the interested
reader is referred to for further details and discussion. 
The general idea is of course to expand about some 
point  $(D_0,d_0)$ on the critical curve  $\E(D_0,d_0)=0$.
The simplest scheme is to extrapolate towards the physical theories
for $D=1,2$ and $d=2,3,\ldots$, using the expansion parameters
$D-D_0$ and $d-d_0$. 
However, as shown in Ref.~\cite{WieseDavid96b}, this set of expansion
parameters is not optimal, and better results are obtained by using 
$D_c(d)=\frac{2d}{4+d}$ and $\E(D,d)=2D -\frac {2-D}2 d$. 
Furthermore, it is advantageous to make expansions for quantities
such as $\nu d$ or $\nu (d+2)$ rather than  $\nu$.

We shall mention three such extrapolation schemes which are based on the
remarkable fact that three different expressions for $\nu$ coincide with
$\nu_0=(2-D)/2$ on the critical line $\E=2D-d\nu_0=0$:
\begin{enumerate}
\item[(1)]  The {\em mean-field} result
\be
\nu_{\mbox{\scriptsize MF}}= \frac{2D}{d} \ ,
\ee
which in the context of polymers and  membranes is known as the 
Gaussian variational approximation \cite{CloJan90,LeDoussal92,Goulian91,GuitterVariational}.
\Eq{final result}  can then be re-organized as
\be
	\nu d =2 D + a(D) \E \ ,
\ee 
leading to an expansion about the mean-field result, which can be plotted
as a function of the expansion point $D$.

\item[(2)] The {\em Flory} expression
\be
\nu_{\mbox{\scriptsize Flory}} =  \frac{2+D}{2+d}\ ,
\ee
can similarly be used as the basis for the expansion of the 
quantity $\nu(d+2)$.

\item[(3)] The {\em large $N$} limit of the theory can be solved exactly,
as will be demonstrated in the next section. The corresponding exact result
\be
\nu_{N\to\infty} = \frac{D}{d-\frac{2D}{2-D}}\ ,
\ee
is also equal to $\nu_0$ for $\E=0$, and can be used as a basis for 
expansion of the quantity $\nu \left( d-\frac{2D}{2-D} \right)$.
\end{enumerate}

After selecting one of these schemes, the next step is to re-express
\Eq{final result} in terms of $D_c(d)=\frac{2d}{4+d}$ and $\E=2D-\frac{2-D}2 d$.
For example, for $\nu (d+2)$, the final result is
\be
\nu(d+2) = 2 +  D_c(d) +\left[  \frac{ 1+{c(D_0)N/2}  }
{\Res\bigg< \GM \bigg | \GB \bigg>  +1 + c(D_0) N/4} -\frac{D_0}2\right] \frac{2+D_0}{2D_0}\E(D,d) \ .
\ee
If we are interested in the  field theory ($D=1$) in 
$d=3$, we have to evaluate the above expression for $\E=1/2$. 
However, we are still free 
to choose the expansion point along the critical curve, which then fixes
$D_0$. As the expansion point is varied, different values for $\nu(d+2)$
(i.e.\ $5\nu$ for $d=3$) are obtained, as plotted in Fig.~\ref{extr1}.
The criterion for selecting a value for $\nu$ from such curves 
is that of minimal sensitivity to the expansion point $D_0$.
We thus evaluate $\nu$ at the extrema of the curves.
The broadness of the extremum provides a measure of the goodness of the result, 
and the expansion scheme. The robustness of this choice in the case of $N=0$
was explicitly checked in Ref.~\cite{WieseDavid96b}, by going to the second order.
(For additional discussions of such ``plateau phenomena'' see Sec.~12.3 of 
Ref.~\cite{WieseDavid96b}.) 

While we examined several such curves, only a selection is reproduced in 
 Figs.~\ref{extr1}, \ref{extr2}, and \ref{4 curves}. We start by checking the method 
for polymers ($N=0$) and the Ising model ($N=1$) in $d=2$, where the exact 
values are known ($\nu=3/4$ and
1 respectively.) The results are given in Table \ref{nutab1}. Only the 
extrapolation for $\nu d$ yields acceptable results; that of $\nu(d+2)$
is bad for $N=1$, and $\nu$ is even worse in both cases. 
 Furthermore, we observe that $c(D)=D$ gives exponents closer to 
the correct value. 
Based on this experience, we focus on the expansions for $\nu d$
in $d=3$, with $c(D)=D$, shown in Fig.~\ref{4 curves} for $N=0$, 1, 2 and 3.
The values of $\nu$ extracted from the maxima are given in Table
\ref{nutab2}, along with their best known estimates from Ref.~\cite{Zinn}. 
Our results are clearly better than the standard 1-loop expansion
of 
\be
\nu = \frac12 + \frac{N+2}{4 (N+8)} \ .
\ee 
There are, however, systematic differences in the trends. In particular, 
we make the observation that the ``exact'' exponents in the range
$0 \le N \le 3$ approximately fall on a straight line with 
slope $0.042 \pm 0.003$; while our results have a perceptible downward
curvature. It may be that the 1-loop results are somehow most suited
to give the exponents at small $N$. A similar suggestion was made in 
Ref.~\cite{BrezinHikami96} in the context of the non-linear sigma model.
Based on these observations, we now pursue an alternative 
expansion, that searches for the slope of $\nu(N)$ at $N=0$. 
 Fig.~\ref{extr2} shows the result for $\partial \nu(N)/\partial N|_{N=0}$
as a function of the expansion point. A very flat and well defined 
plateau is obtained, with a value of 0.042 in excellent agreement
with the slope quoted earlier. Of course, to get the absolute value 
of the exponents, we also need to specify $\nu(N=0)$. As discussed in 
Ref.~\cite{WieseDavid96b}, the best expansion quantity for this purpose
is $\nu(d+2)$ leading to $\nu(N=0)=0.59$ at the 1-loop order. (This exponent
is also obtained if one demands  the 1- and 2-loop results to be equal.)

\section{The limit $N\to \infty$ and other approximations}
\label{The limit N to infty and other approximations}
As in the $O(N)$ model, it is possible to derive the dominant 
behavior for large $N$ {\em exactly}.
In the  standard $\varphi^4$-theory,
one starting point is the observation that 
\be \label{cond N to inf}
\left<(\vec S^2)^2(r) \right>=\left< \vec S^2(r) \right>^2  \ ,
\ee
since in the limit $N\to \infty$, spin-components of different colors decouple
\cite{Zinn,MAInDomb,Amit}.
This is also known as the random phase approximation (RPA).

Here, we pursue a slightly different approach, based on the diagrammatic
expansion. Note that for $N\to \infty$, 
 only simply connected configurations survive.
(The vertices are made out of membranes, the links out of 
$\tilde \delta^d$-interactions.) 
For example, the diagram  $\,\PEclose\,$ which is doubly connected,
and the diagram $\,\PFclose\,$ which includes a self-interaction,
each have one factor of $N$ less than the simply connected graph $\,\PCclose\,$.
 The leading diagrams for the membrane density at the origin are then given by
\vspace{-7mm}
\be
f:=\left< \int_{x} \tilde \delta^d(r(x)) \right> = \,\,\PA\,+\,\,\PB\,+\,\PJ\,+\,\PK\, + \ldots\ .
\ee
The above sum can be converted into a self-consistent equation for $f$
by noting the following: Successive diagrams can be obtained from the
first (bare) diagram by adding to each point of a manifold 
 a structure that is equivalent to $f$ itself.
This is equivalent to working with a {\em single non-interacting manifold}
for which the chemical potential $t_0$ is replaced by an effective
value of $t_0+b_0f$.
Calculation of $f$ for this manifold proceeds exactly as in \Eq{Z1(0)mem},
and results in the integral
\be \label{rel f}
f=\int \frac{\rmd \Omega}{\Omega} \Omega^{1-\nu_0d/D} \rme^{-\Omega (t_0+b_0 f)} 
\ .
\ee
The above integral is strongly UV-divergent, and leads to a form
\be \label{rel f'}
f = B (t_0+b_0 f)^{\frac{2-D}{2D}d -1} + A \ , 
\ee
where 
\be
B= \Gamma\left(1-\frac{2-D}{2D}d \right)\ .
\ee
The strong UV-divergence, controlled with an explicit UV-cutoff, 
is absorbed in the constant $A$. 
It is usually dropped in a dimensional regularization scheme, as in \Eq{Z1(0)mem}.

The radius   of gyration $R$ is now related to $f$ as follows:
From \Eq{rel f} we note that $t_0 + b_0 f$ is the physical chemical potential 
conjugate to $\Omega$, thus leading to a typical volume 
of $\Omega\sim 1/(t_0 + b_0 f)$.
Since there are no self-interactions in the effective manifold introduced above,
its radius can be related to the volume by $R\sim \Omega^{\nu_0/D}$.
Thus, up to  a numerical factor which is absorbed into the definition of $R$, we obtain
\be \label{rel 2}
R^{-\frac{2D}{2-D}} = t_0+ b_0 f \ .
\ee
Eliminating $f$ in  \Eq{rel f'} with the help of \Eq{rel 2} yields
\be \label{rel3-}
R^{-\frac{2D}{2-D}} =  (t_0+ b_0 A) +  b_0 B \,
R^{\frac{2D}{2-D} - d}
\ . 
\ee
Identifying the difference in temperature to the critical theory as
\be \label{rel3}
\bar t = t_0 +  b_0  A\ ,
\ee
the critical theory is approached upon taking $\bar t \to 0$ and $R\to \infty$. 
This occurs if and only if  $d$ is larger than the lower critical dimension
\be
d>d_{l} = \frac {2D}{2-D} \ .
\ee
If $d$ is in addition smaller than the upper critical dimension, i.e.
\be
d<d_u=\frac{4D}{2-D}\ ,
\ee
the left hand side of \Eq{rel3-} vanishes faster than the 
$R$-dependent term on the right hand side, and we obtain the scaling relation
\be
R \sim {\bar t}^{\,-1/(d-\frac{2D}{2-D})} \ .
\ee
In the large $N$ limit, the exponent $\nu^*$ is therefore given by
\be \label{nu large N}
\nu^*_{N\to\infty} =\frac D{d-\frac{2D}{2-D}}
\ .
\ee
We can verify that the standard result\cite{Zinn} is correctly reproduced for $D=1$ as
\be 
\nu^*_{N\to\infty}(D=1) = \frac1{d-2} \ .
\ee 
Note that for  $d>d_u$, the leading behavior from \Eq{rel3-} is
\be
R\sim \bar t^{\,-\frac{2-D}{2D}}\ , 
\ee
implying the free theory result
\be
\nu_0=\frac{2-D}{2} \ .
\ee

It is interesting to cast the other approximation schemes introduced in the 
previous section in the language of renormalization factors. 
The Flory-approximation assumes that the elastic energy and the contribution due
to self-avoidance scale in the same way. This enforces for the 
renormalization factors $Z$ and $Z_b$ (in the massless scheme with $Z_t=1$)
the constraint
\be \label{Z for Flory approximation}
	Z=Z_b \ .
\ee
Knowing such a relation, the critical exponent $\nu$ can be calculated
explicitly \cite{WieseDavid96b}: Suppose that
\be \label{8.14}
Z_b= Z^\sigma\ , \quad\  \mbox{and} \qquad Z_t=1 \ .
\ee
From the definition of the  $\beta$-function in \Eq{RG-functions}, 
we obtain 
\be \label{8.15}
\beta(b) \frac{\partial}{\partial b} \ln (Z_b Z^{d/2} ) = -\E -\frac{\beta(b)}{b} \ .
\ee
The second term on the r.h.s.\ can be neglected upon approaching the 
critical point $b^*$, where the $\beta$-function vanishes.
Inserting \Eq{8.14} into \Eq{8.15}, solving for $\beta(b) \frac{\partial}{\partial b} \ln Z$, and substituting the result into \Eq{RG-functions'} yields
\be
\nu^*= \frac{2-D}{2} + \frac{\E}{d+2\sigma} \ .
\ee 
 For the Flory-approximation, $\sigma=1$ from \Eq{Z for Flory approximation}, 
and the above expression evaluates to
\be
\label{nu Flory}
\nu_{\mbox{\scriptsize Flory}} = \frac{2+D}{2+d}
\ .
\ee

The last approximation scheme that we shall discuss is a mean-field limit. 
As is well known from the mean-field approximation to $\varphi^4$-theory, 
minimizing an expansion of the order parameter (e.g. by a saddle point method)
results in an exponent  $\alpha=0$ describing the singularity in the heat capacity.
Assuming that the singular part of the free energy scales as 
\be
f_{\mbox{\scr sing}}\sim R^{-d}\sim \Omega^{-d\nu^*/D}\sim
\overline{t}^{\,d\nu^*/D}\ ,
\ee
 leads to a generalized heat capacity exponent for
manifolds given by
\be
\alpha = 2 -\frac{\nu^* d}{D} \ . 
\ee
A discontinuous (but non-diverging) heat capacity then leads to
\be \label{nu mean field}
\nu_{\mbox{\scriptsize MF}} = \frac{2D}d \ ,
\ee
which coincides with the result obtained by considering large $d$ and $N=0$
\cite{LeDoussal92,Goulian91,GuitterVariational}.
In such a limit, and in a massless scheme, corrections due to self-avoidance 
are strongly suppressed \cite{WieseDavid96b}, and we have
\be
Z_b=1 \ ,\qquad  Z_t=1 \ , \qquad Z\ne 1\ .
\ee
This is equivalent to $\sigma=0$ from Eqs.~\eq{8.14} and \eq{8.15},
and again yields \Eq{nu mean field}.

\section{Low temperature expansions of the Ising model}
\label{Low temperature expansions of the Ising model}
In Sec.~\ref{The O(N)-model in the high-temperature expansion}
we demonstrated that the high temperature expansion of 
the $O(N)$ spin model naturally leads to a sum over $N$-colored
loops ($D=1$); motivating the later generalization to manifolds (arbitrary $D$).
 For the Ising model ($N=1$), a related description can be obtained from a low 
temperature expansion. Excitations to the uniform (up or down pointing) ground
state are in the from of droplets of spins of opposite sign. The energy
cost of each droplet is proportional to its 
boundary, i.e.\ again weighted by a Boltzmann factor of the form
$$
\rme^{-t\Omega} \ .
$$
Thus a  low temperature representation of the $d$-dimensional Ising partition function
is obtained by summing over all closed surfaces of dimension $D=d-1$. 
 For $d=2$, the high and low temperature series are similar, indicating 
the self-dual nature of the model. For $d=3$, the low temperature description
is a sum over surfaces, which is also the high temperature expansion of an 
Ising lattice gauge theory \cite{Savit1980}, establishing the duality between these
two models. 

The non-trivial question is regarding the type of surfaces which dominate the above
sum. There is certainly no constraint on the internal metric, in contrast to the 
{\em tethered surfaces} in $D=2$ which have a flat metric. 
Since the sum includes droplets of all shapes, it may be more appropriate to
examine  {\em fluid membranes}. However, there is currently no
practical scheme for treating interacting fluid  membranes, and the
excluded volume interactions between the membranes 
are an essential ingredient to avoid overcounting configurations. 
We shall argue that, at least in low dimensions, the sum over tethered 
membranes captures the appropriate physics of the problem. 
This may appear quite surprising at first glance, as for $N=0$, 
surfaces generated from pla\-quettes
on a lattice are very different from tethered surfaces.
The former are dominated by configurations that resemble branched polymers. The 
large entropy gain of branches is responsible for this, and appears as an instability
towards formation of spikes in a string theory \cite{AmbjornLH94}. 
However, it is possible that for $N>0$, the above instability is replaced
by a string of bubbles. (Reminiscent of the Raleigh instability of a stream in
hydrodynamics.) The collection of bubbles is then satisfactorily described by
a set of fluctuating hyperspherical (tethered) manifolds which is the basic ingredient
of our model. The appropriate question may be whether tethered membranes sweep 
out phase space, i.e.\ form a complete set of basis functions in the
configuration space of our problem.

Another issue is whether the sum may be restricted to spheres, or if
 objects of 
other topologies must also be included. We argue in Appendix \ref{Other topologies}
that the dominant contribution (and the only one included in perturbation theory)
is the one from spheres.

\begin{figure}[htb]\centerline{
\epsfxsize=0.5\textwidth \parbox{0.5\textwidth}{\epsfbox{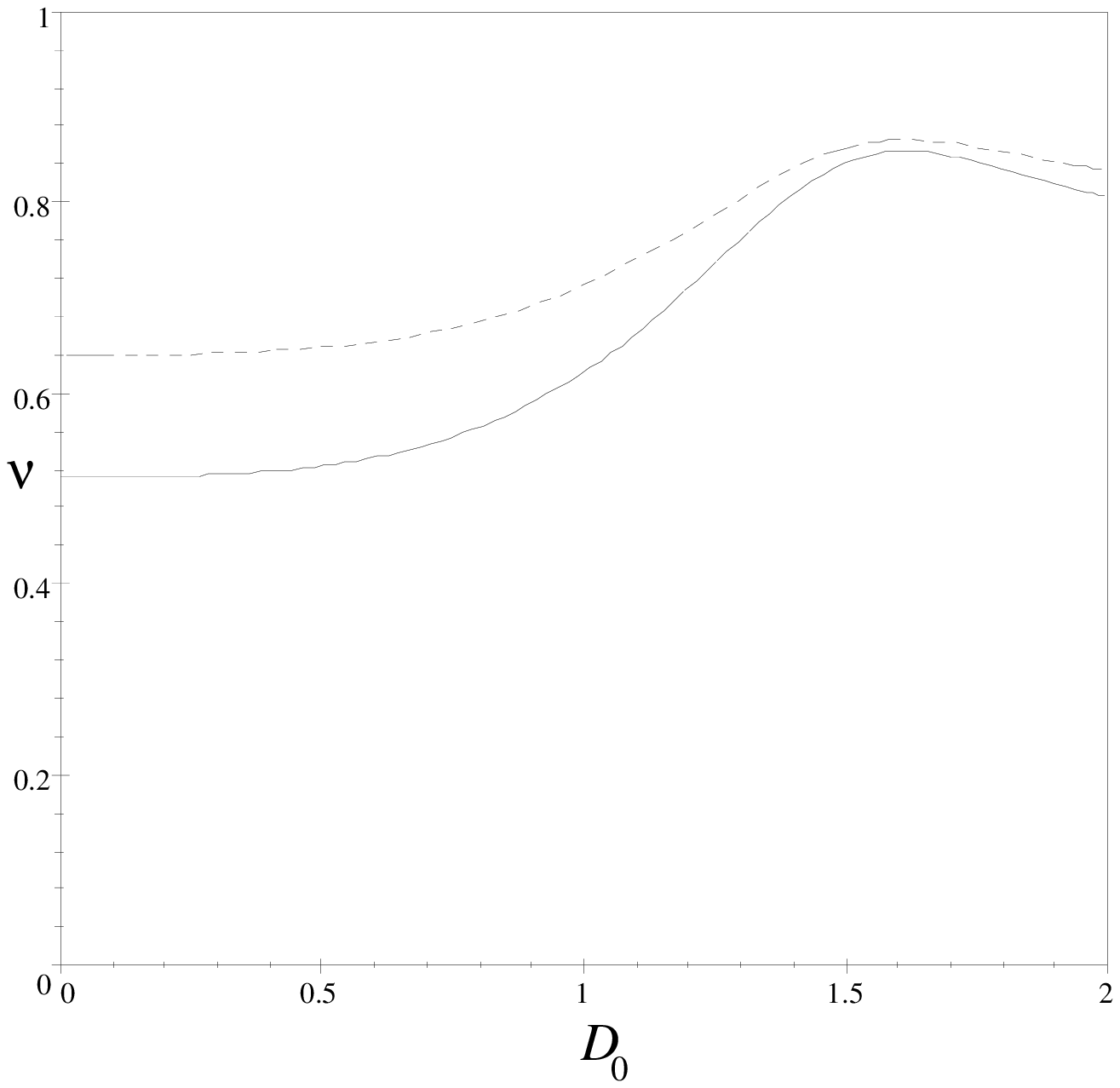}}
\epsfxsize=0.5\textwidth \parbox{0.5\textwidth}{\epsfbox{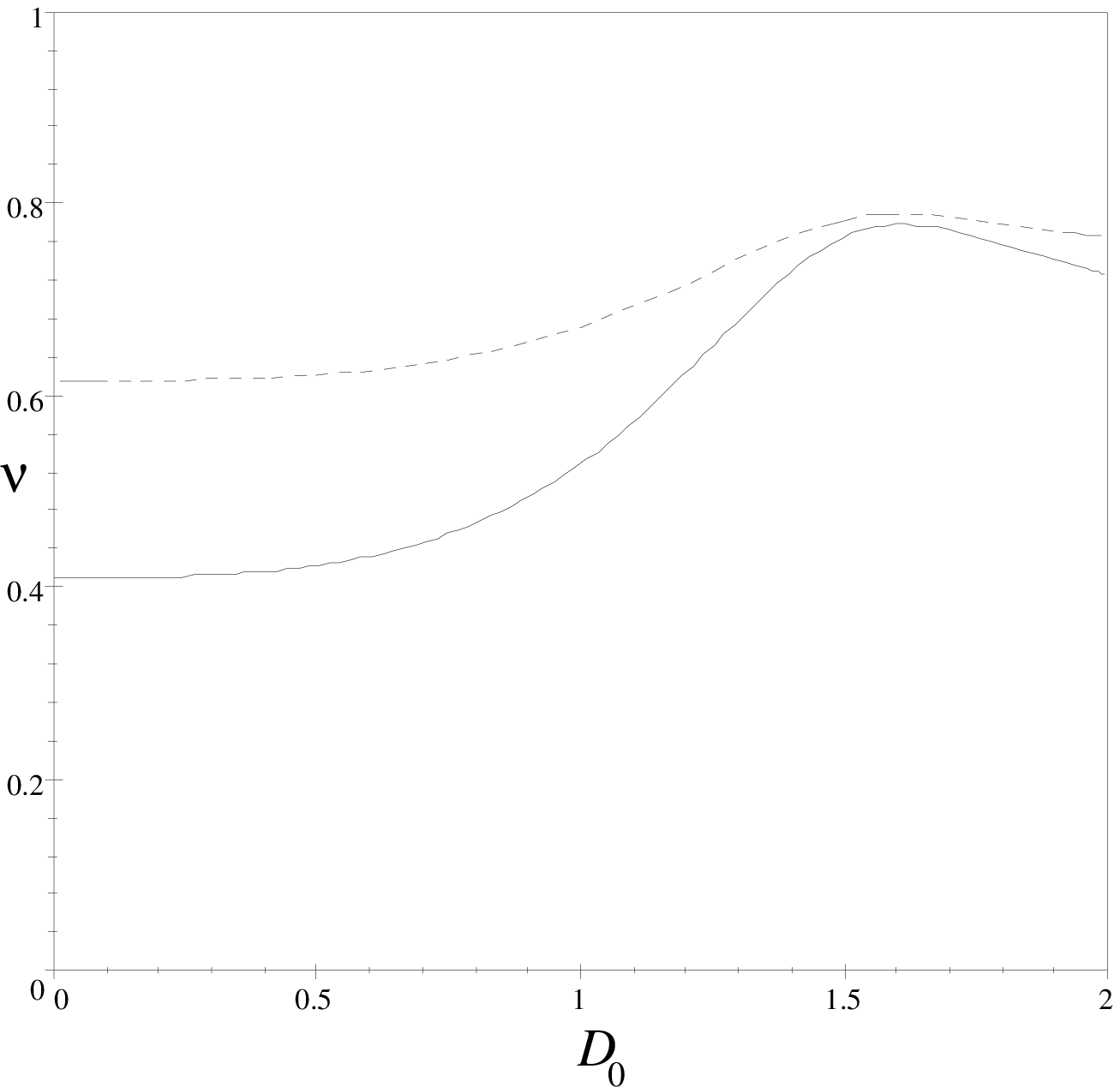}}}
\centerline{
\epsfxsize=0.5\textwidth \parbox{0.5\textwidth}{\epsfbox{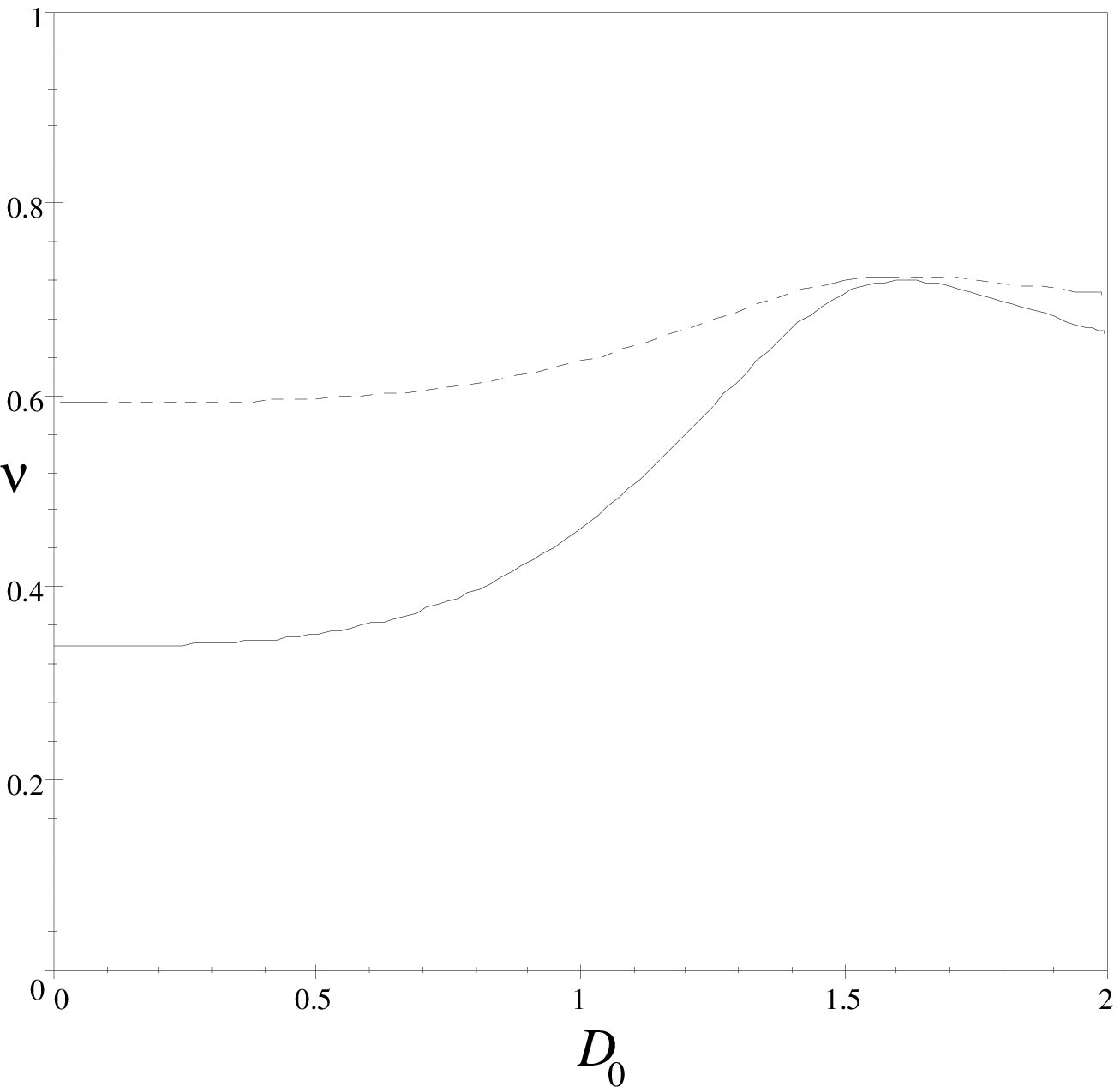}}
\epsfxsize=0.5\textwidth \parbox{0.5\textwidth}{\epsfbox{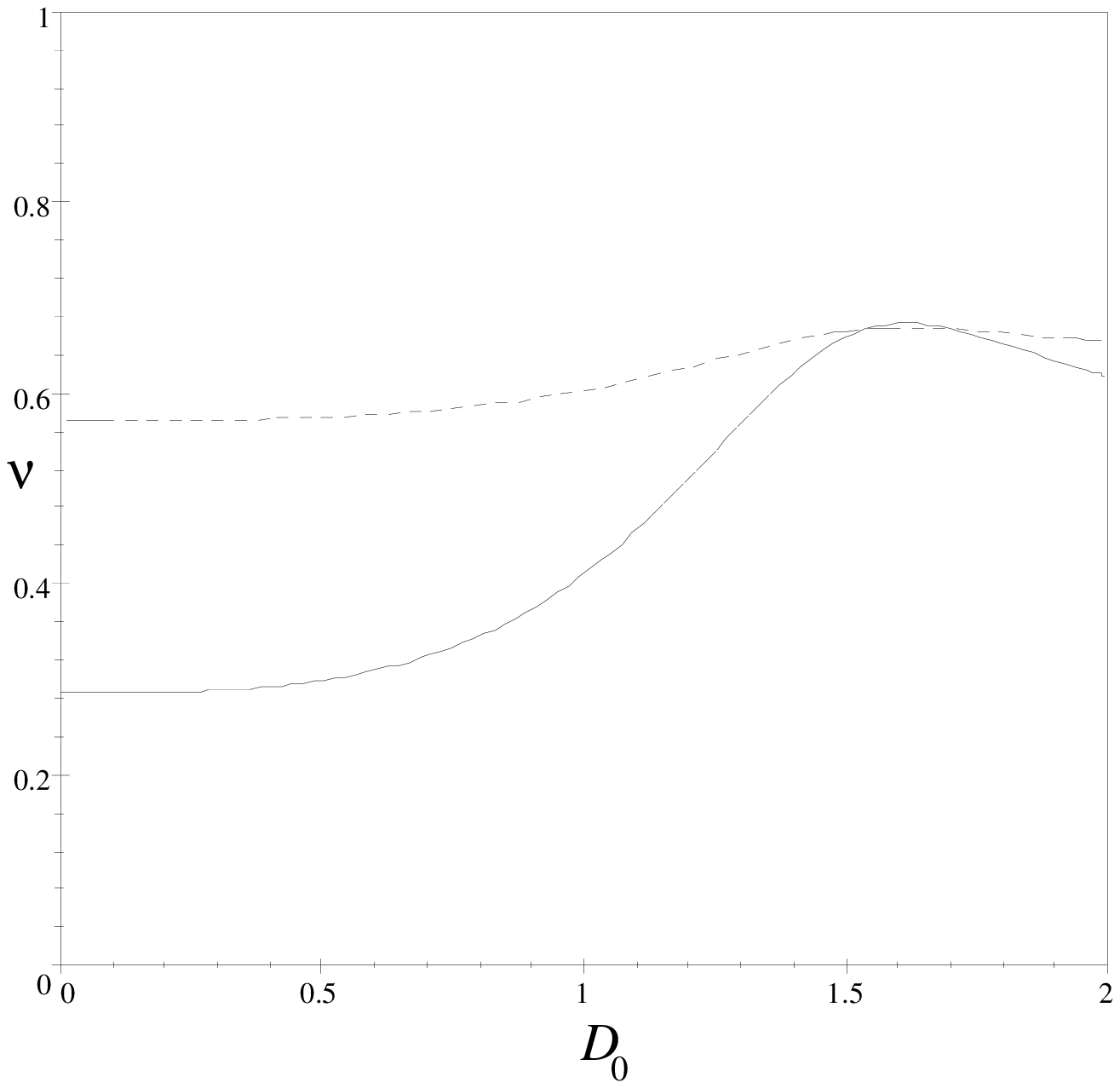}}
}\caption{Test of the relation in \Eq{nu hypoth1} for the Ising-model
in $d=2.25,\ d=2.5,\ d=2.75$ and $d=3$. The upper curves are from the
high temperature representations of the Ising model ($D=1$), 
while the lower curves are from the low temperature expansion ($D=d-1$)
as explained in the text.  The exponent $\nu$ is extrapolated from the 
maximum of each curve. 
These curves are based on the scheme in which we extrapolate $\nu d$, 
linearize in $N$,  and to divide the result by $dD$. 
This choice is determined through the quality
checks of section \protect\ref{Extrapolations}.}
\label{specIsing}
\end{figure}%
We shall now test the validity of the above conjecture. As discussed before,
our generalization to the sum over manifolds is defined only up to
the factor $c(D)$.  For the remainder of this section we make the 
``least number of variables'' choice $c(D)=D$, for the partition function. 
Singularities of the partition function are characterized by the critical
exponent $\alpha(D,d,N)$, through
\be
	\ln {\cal Z}_{\mbox{\scr singular}} \sim |t-t_c|^{2-\alpha} \ .
\ee
The equality of the singularities on approaching the critical point from 
the low or high temperature sides, requires
\be
	\alpha(1,d,1) = \alpha( d-1,d, 1)
\ .
\ee
 From the scaling-relation
\be
\alpha(D,d,1) = 2- \frac{\nu( D,d, 1)d}D
\ee
we obtain
\be \label{nu hypoth1}
{\nu(1,d,1)} =  \frac{1}{d-1}\,\nu(d-1,d, 1)
\ .
\ee
A more general identity is obtained by considering a generalized 
class of Ising models, whose interactions are defined on $D$-dimensional 
primitive elements of a lattice  \cite{Savit1980}.
In the standard Ising model the interactions occur on {\it bonds}
($D=1$), while in Ising lattice gauge theory interactions
are placed on plaquettes ($D=2$). 
Equating the singularities from the high and low temperature expansions of 
such models, {\em and assuming a single continuous phase transition}, yields
\be \label{nu hypoth}
{\nu(D,d,1)} =  \frac{D}{d-D}\,\nu(d-D,d, 1)
\ .
\ee

The conjectured identity in \Eq{nu hypoth1} was tested numerically, and the results
are presented in Fig.~\ref{specIsing}. 
The extrapolated exponents (the maxima of the curves) from the dual high and
low temperature expansions are in excellent agreement. Indeed, one could hardly
expect better from a 1-loop calculation.
Nevertheless, higher-loop calculations would be useful to check
this surprising hypothesis. 
One of the peculiarities of the 1-loop extrapolations presented in 
Fig.~\ref{specIsing} 
is that any crossing of the curves from the dual models occurs
at the {\em mean--field} value of $2D/d$ from \Eq{nu mean field}.
This is accidental, but present for all of our ``good'' extrapolation
variables. An explicit counter example is given on Fig.~\ref{specIsing2},
where we used the ``bad'' extrapolation-variables $1/D$ and $d$. 
They are bad, as there is no pronounced  maximum to estimate  $\nu$.
Nevertheless, the intersection of the two dual models occurs at 
$\nu=0.6$, not too far from the exact result of 0.6315.
\begin{figure}[t]\centerline{
\epsfxsize=0.7\textwidth \parbox{0.7\textwidth}{\epsfbox{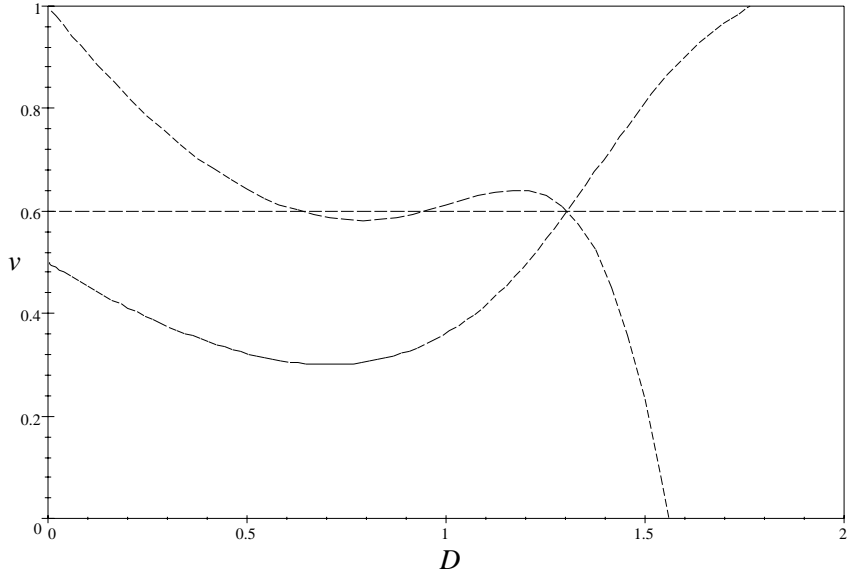}}
}
\caption{Extrapolations for the Ising model and its dual in $D=1$, 
$d=3$, using $c(D)=D$, and  as extrapolation variables $1/D$ and $d$.}
\label{specIsing2}
\end{figure}

\section{Cubic anisotropy}
\label{Cubic anisotropy}
Up to now, we assumed that the interaction between 
manifolds is independent of their color. 
This is a consequence of the rotational invariance of the underlying spin
model introduced in Sec.~\ref{The O(N)-model in the high-temperature expansion}. 
This equality of interactions does not have to
hold in a system of polymer loops. In the context of the $\varphi^4$-theory,
unequal interactions result from the breaking of rotational symmetry, 
e.g.\ through the introduction of cubic anisotropy, 
as discussed in Refs.~\cite{Amit,KetWal73,Aharony73,BrezinGillouZinn74}. 
In the microscopic spin model of Sec.~\ref{The O(N)-model in the high-temperature expansion}, the independence of the interactions
between loops from their colors emerges as a consequence of the normalization 
condition
\be
\sum_i S_i^2 =1 \ .
\ee
If we replace this by the constraint
\be
\sum_i |S_i|^a =1 \ ,
\ee
with $a\ne 2$, which breaks rotational symmetry, this is no longer the case. 

The model in the absence of full spherical symmetry is described by
two interaction parameters. In addition to $b$ which indicates the
interaction  between any two membranes irrespective of their color,
there is a new anisotropic  coupling constant $u$
which acts only between membranes of the same color. 
Physically, not all combinations of $u$ and $b$ are allowed, as
some of them induce  a collapse of the  membranes.
The following two cases can be distinguished: 
\begin{figure}[t]\centerline{
\epsfxsize=0.5\textwidth \parbox{0.5\textwidth}{\epsfbox{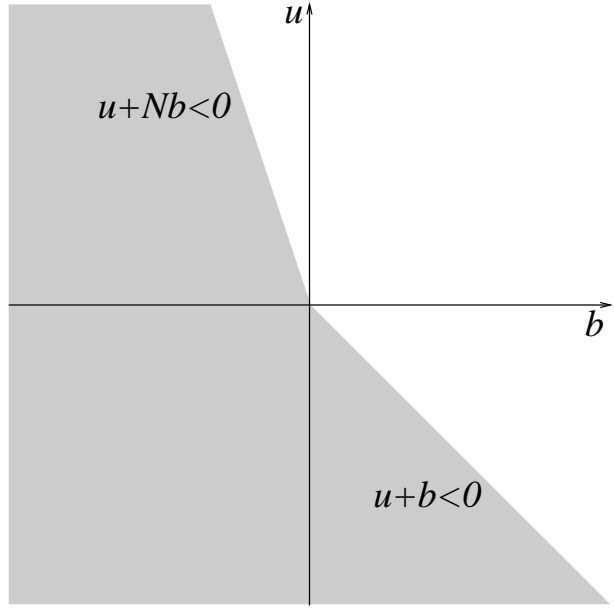}}
}
\caption{Regions of stability for the bare model with cubic anisotropy
(the white portions are stable): 
The bare model is unstable if  $u+Nb<0$ for $u>0$, and if $u+b<0$ 
for $u<0$. Note that this diagram changes upon renormalization.}
\label{stability}
\end{figure}

{\bf (i)} If $u$ is negative, for a single membrane to avoid itself and
not to collapse, the condition
\be \label{u<0}
u+b>0 \ ,
\ee
has to hold. This implies that $b$ is positive, and the repulsive  interactions
between any pair of membranes ensures mutual stability.
The same condition is obtained  in the $O(N)$-model by requiring  the 
stability of the minimum energy state \cite{Amit}.

{\bf (ii)} If $u$ is positive, the stability condition in the $O(N)$-model
is \cite{Amit}
\be \label{u>0}
u+Nb>0 \ .
\ee
This places a lower bound on $b$ which is, for more than one color ($N>1$),
more restrictive than \Eq{u<0}, but still admits negative values for $b$ (see 
Fig.~\ref{stability}). For $b<0$, membranes of different color attract;
in the extreme limit becoming glued together to form
a ``super-membrane'' out of  $N$ differently colored membranes. 
In this limit, the theory reduces to an Ising-like system, where the effective number
of colors is one.  (The corresponding RG-flows, as discussed below, do
indeed tend to an Ising fixed point.) For this ``super-membrane''
not to collapse, we again find the condition in \Eq{u>0}. 
However, this picture is quite schematic, and real physical systems are
governed by many more parameters, and may well behave differently. 
Studies in polymers  \cite{SchaeferPrivate,Ebert96} indicate that
the precise competition of the attractive and repulsive parts of the
interaction potential plays a crucial role, sometimes leading to  non-universal 
behavior.  

The above stability arguments are based on energetic considerations, and 
are expected to be modified upon the inclusion of fluctuations, say through
a renormalization group procedure.
In the studies of critical phenomena, a well-known example is the
Coleman-Weinberg mechanism \cite{Amit,ColemanWeinberg1973},
where the RG flows take an apparently stable combination of $b$ and
$u$ into an unstable regime, indicating that fluctuations destabilize the system.
In the flow diagrams described below, we also find the reverse behavior
in which an apparently unstable combination of $b$ and $u$ flows to
a stable fixed point. We interpret this behavior as indicating that fluctuations
actually stabilize the model, a reverse Coleman-Weinberg effect,
which to our knowledge has not been discussed before.
To decide whether a system with a given combination of couplings
is stable, we first follow the renormalization group flow.
As perturbative renormalization does not say anything about the
strong coupling regime, we regard run-away-trajectories in the flow 
as indicating an unphysical situation. 
If, on the other hand the renormalization group flow tends to a finite
and completely IR-stable fixed point, we use the ``classic''
stability analysis discussed above,
since renormalization has eliminated {\em all} fluctuations.
Using this criterion we have shaded in 
grey the unphysical regions in the following flow diagrams.

The derivation of the generalized renormalization group functions 
is most easily carried out in a mixed scheme, in which we absorb 
contributions due to self-avoidance  into $Z$, and those due to 
interactions with another membrane (proportional to $N$) into $Z_t$.
For the original model, this scheme leads to the renormalization factors 
\bea
Z_t(b)&=& 1+\frac b{2\E} \frac{N c(D)}2  \ ,\nn\\
Z(b)&=&1+\frac{2-D}{2D} \frac b \E \ .
\eea
The presence of the additional interaction $u$ between membranes of the
same color, modifies the above result to  
\bea \label{new Z's}
Z_t(b,u)&=& 1+ \frac{b N c(D)}{4\E} + \frac{uc(D)}{4\E} \ , \nn\\
Z(b,u)&=&1+\frac{2-D}{2D} \frac {b+u} \E \ .
\eea

To derive the renormalization of the 
coupling-constants, we note that with 
this choice of $Z$ and $Z_t$, we have eliminated all divergent 
configurations which in the polymer picture and in the MOPE are 
denoted by
\be
\PF\ ,\quad \mbox{ and}\qquad \FD \ ,
\ee
respectively. 
The diagrams that renormalize the interactions are of two classes
\be
\PE\ , \qquad \mbox{\rm and}\qquad \PC \ .
\ee
In calculating the contribution to the renormalization of $b$, the inner
loop of the latter diagram contributes a factor of $N$ if both interactions
are $b$, and 1 for the two cases when one of the interactions is $u$,
resulting in
\bea
b&=&\mu^{-\E } Z^{-\frac{d}{2}} 
\left( 1-\frac b\E \Res \bigg< \!\GM \bigg | \GB\! \bigg>
-b \frac{Nc(D)}{4\E} -u \frac{c(D)}{2\E}\right)b_0 
\ .\quad
\eea
(Note that because of the choice of the mixed RG scheme, the above 
relation does not
reduce for $u=0$ to the corresponding one derived earlier.)
For the anisotropic coupling $u$, the first class of diagrams can be constructed
from two interactions $u$ (1 way), or one $u$ and one $b$ (two ways).
The second class gives a single contribution proportional to $u^2$,
for the overall result
\bea 
u&=&\mu^{-\E} Z^{-\frac d2} 
\times \left( 1 -\frac {u+2b}\E \Res \bigg< \GM \bigg | \GB \bigg>
-u \frac{c(D)}{4\E} \right)u_0 
\ .
\eea
The renormalized parameters yield the $\beta$-functions at 1-loop order, as
\bea \label{betab}
	\beta_b(b,u) &=&
 -\E b + b^2 \left(\Res \bigg< \GM \bigg | \GB\bigg>  + 
1 + \frac{Nc(D) }4\right)
+\,bu \left(1+\frac {c(D)} 2 \right) \ , \nn\\&&
\eea
and 
\bea \label{betau}
\beta_u(b,u) &=&
 -\E u + u^2 \left(\Res \bigg< \GM \bigg | \GB\bigg>  + 
1 + \frac{c(D)}4\right) \nn\\
& & 
+\,bu \left(1+ 2 \Res \bigg< \GM \bigg | \GB\bigg> 
\right) \ .
\eea
 Finally, the exponent $\nu$ is obtained from \Eq{new Z's} as
\be
\nu(b,u) =  \frac{2-D}{2} \left( 1+\frac1{2D}\left[ b\left(1+\frac{N c(D)}{2}
\right)+u\left( 1+\frac {c(D)} 2 \right)\right] \right)
\ .
\ee
 For $D\to 1$, these equations reduce to the renormalization-group functions 
reproduced in Amit \cite{Amit}. (In \cite{Amit} there is
 an additional factor of 2/3 due to the choice of numerical constants.) 
 
As in their standard counterpart, these flow-equations admit 4 fixed points:
\begin{enumerate}
\item[(1)] The {\em Gaussian} fixed point
\be
b^*_G=0 \ ,\quad u^*_G=0 \ ,
\ee 
which is always unstable below the upper critical line $d_u(D)$.
(The following discussions of stability all pertain to this region.)

\item[(2)] The {\em Heisenberg} fixed point
\be
b^*_H=\frac{\E}{1+\frac{Nc(D)}4+\Res \bigg< \GM \bigg | \GB\bigg>} \ ,
 \quad u^*_H=0\ ,
\ee
is always stable along the $u=0$ axis, and is completely stable
as long as 
\be
\frac{N c(D)}{4} < 	\Res \bigg< \GM \bigg | \GB\bigg>_{\E^{-1}} \ .
\ee 
This certainly applies to single polymers and membranes with $N=0$, 
but is also the case for the $O(N)$ model as long as $N<4$. 
\item[(3)] The {\em Ising} fixed point
\be
b^*_I=0\ ,\quad u^*_I=\frac{\E}{1+\frac{c(D)}4+\Res \bigg< \GM \bigg | \GB\bigg>} \ ,
\ee
is stable if 
\be
\Res \bigg< \GM \bigg | \GB\bigg>< \frac{c(D)}{4} \ .
\ee
For the standard $O(N)$-model with $D=1$ the Ising fixed point is unstable.
However, since for large $d_c$, the left hand side of the above inequality
decays rapidly as $2^{-d_c/2}$, the Ising fixed point is stable if the expansion
point is sufficiently close to $D=2$. 

\item[(4)] The {\em cubic} fixed point is located at
\two{\end{multicols}\widetext\noindent\lineup}{}
\bea
\two{}{\hspace{-7mm}}b^*_c  &=& 
\frac{\left( A(D)-\frac{c(D)}{4}\right)
}{\left(1+\frac{Nc(D)}4+ A(D)\right)
\left( A(D)-\frac{c(D)}{4}\right)+
\left( 1+\frac{c(D)} 2 \right) 
\left(\frac{Nc(D)}4 -  A(D)\right)
}\E \ ,\nn\\
\two{}{\hspace{-7mm}}u^*_c  &=& 
\frac{
\left(€\frac{Nc(D)}{4} -  A(D)\right)
}{\left(1+\frac{Nc(D)}4+ A(D)\right)
\left( A(D)-\frac{c(D)}{4}\right)+
\left( 1+\frac{c(D)} 2 \right) 
\left(\frac{Nc(D)}4 - A(D)\right)}\E\ ,\nn\\
& &\mbox{where}\qquad A(D)=\Res \bigg< \GM \bigg | \GB\bigg> \ . \label{cubic FP}
\two{}{}
\eea
\two{\begin{multicols}{2}\narrowtext\noindent\linedown}{}
The stability of this point under RG flows depends on the parameters 
$D$ and $N$, and this dependence is not simple. 
The physical stability of this fixed point according to the criteria 
of \Eqs{u<0}  and \eq{u>0} should also be verified.
Interestingly, we find (at least at 1-loop order) that whenever the
fixed point is IR-stable, it falls in the physically stable region.
Note that there are combinations of $D$ and $N$, for which  the cubic fixed 
point is at infinity in the 1-loop approximation. 
An explicit example is for $D=1$ and $N=0$, a scenario relevant to the 
random bond Ising model discussed in the next section.
\end{enumerate}

\begin{figure}[t]\centerline{
\epsfxsize=\textwidth \parbox{\textwidth}{\epsfbox{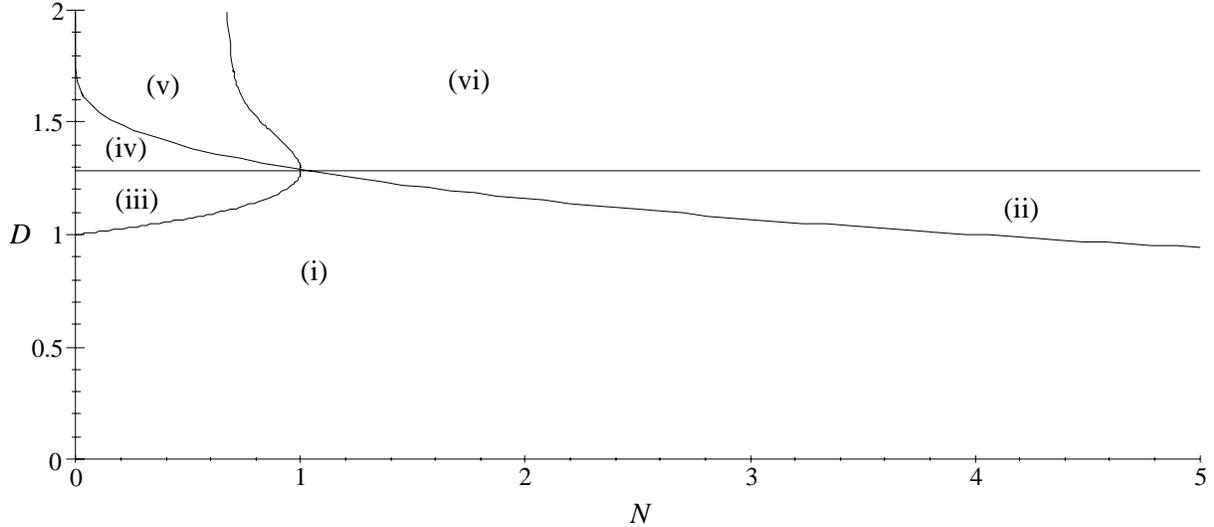}}
}
\caption{Regions with different RG flow patterns in the $(N,D)$-plane, 
as discussed in the text.
The domain-boundaries are obtained from $c(D)=4 A(D)$ for the line separating
(iii) and (iv) as well as (ii) and (vi); $N c(D)=4 A(D)$ for the line 
separating (i) and (ii) as well as (iv) and (v). A third boundary is 
given by the vanishing of the denominator in \Eq{cubic FP}. }
\label{domains}
\end{figure}%

The two RG equations admit six different flow patterns, the last four 
of which do not occur in the standard field theory. 
The domains of the $(N,D)$ plane corresponding to each of the 
following scenarios is plotted  in Fig.~\ref{domains}:
\begin{enumerate}
\item[(i)] For $D=1$, i.e.\ in the case of the $O(N)$ model, and for $0<N<4$, 
only the Heisenberg fixed point is stable, as indicated in Fig.~\ref{flow1}. 
This is the fixed point that is usually studied in the context of critical phenomena. 

\item[(ii)] For $D=1$ and $N>4$, the Heisenberg fixed point is unstable and the 
system is governed by the cubic fixed point (Fig.~\ref{flow2}).

\item[(iii)] An interesting phase diagram is obtained for $N=0$ and $1<D<1.29$. 
Then, as shown in Fig.~\ref{flow5}, the Heisenberg and cubic fixed points
are {\em both} stable, their domains of attraction being separated by the
axis $b=0$.

\item[(iv)] Another phase diagram with two completely stable fixed points
is obtained for $N=0$ and $D$ close to 2. 
Then, as shown in Fig.~\ref{flow3}, the Heisenberg and Ising fixed points
are both stable, and there is a phase separatrix passing through the 
Gaussian and the cubic fixed points.

\item[(v)]
For $D$ close to 2 and $N>N_c$, where $N_c$ vanishes exponentially for
$D\to 2$, the cubic fixed point is in the lower right sector.
Both the cubic and Ising fixed points are stable as indicated in Fig.~\ref{flow6}. 

\item[(vi)] Yet another possibility is that both the Heisenberg and the 
cubic fixed points are unstable, as in  Fig.~\ref{flow4}. This is the case 
for $D$ close to 2 and $N$ large.
 Then only the Ising fixed point is 
attractive and controls the critical behavior.
\end{enumerate}%
\begin{figure}[t]\centerline{
\epsfxsize=0.5\textwidth \parbox{0.5\textwidth}{\epsfbox{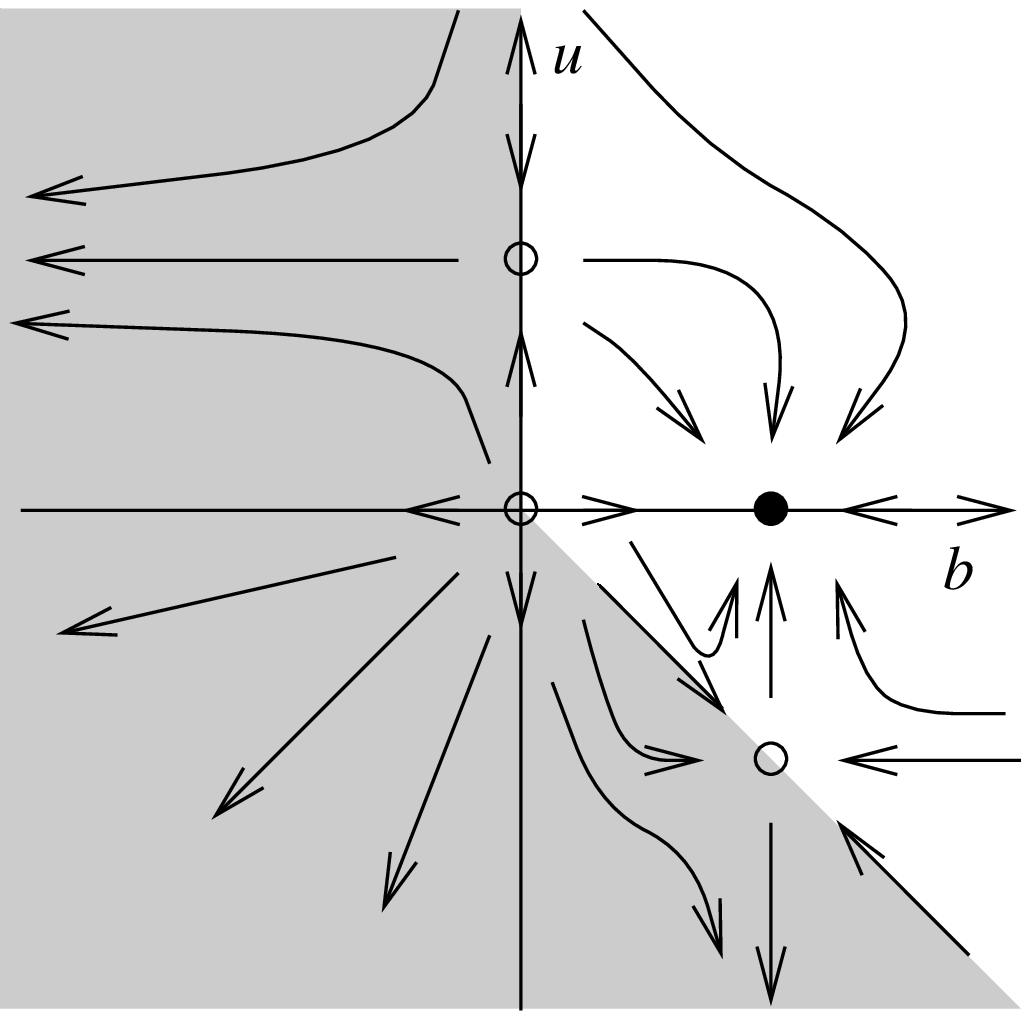}}
}
\caption{RG-flow from Eqs.~\eq{betab} and \eq{betau} in domain (i), e.g.\ $D=1$ and $N<4$;
shaded regions are unstable.}
\label{flow1}
\end{figure}%
\begin{figure}[t]\centerline{
\epsfxsize=0.5\textwidth \parbox{0.5\textwidth}{\epsfbox{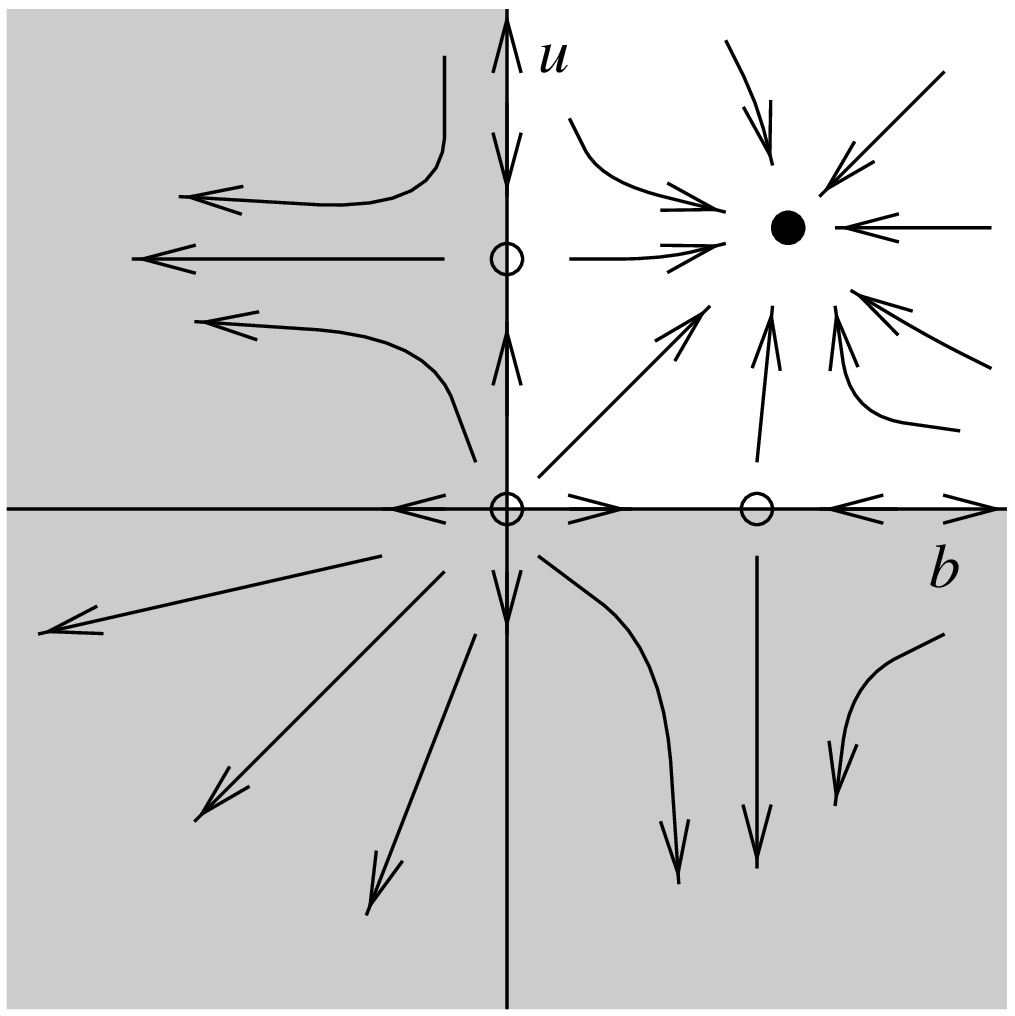}}
}\caption{RG-flow from Eqs.~\eq{betab} and \eq{betau} in domain (ii), e.g.\ $D=1$ and $N>4$;
shaded regions are unstable.}
\label{flow2}
\end{figure}%
\begin{figure}[t]\centerline{
\epsfxsize=0.5\textwidth \parbox{0.5\textwidth}{\epsfbox{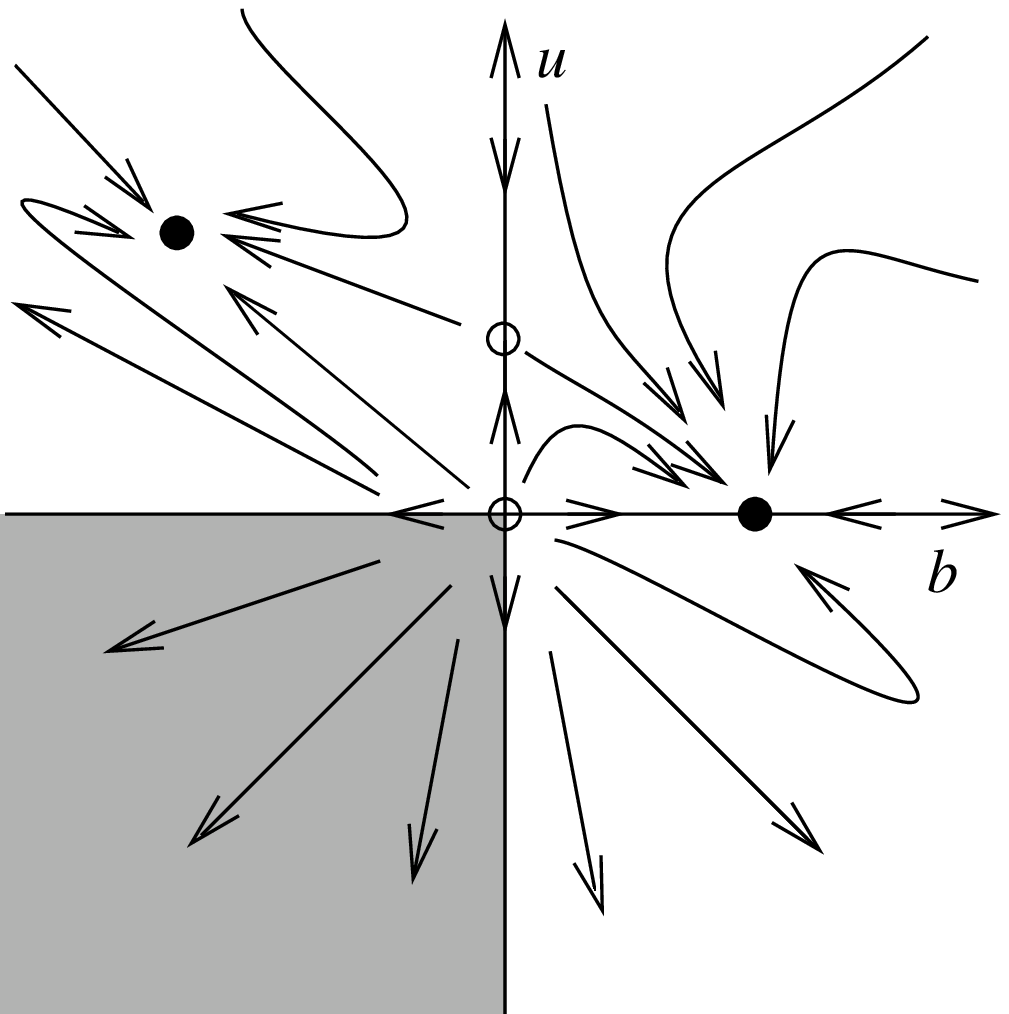}}
}\caption{RG-flow from Eqs.~\eq{betab} and \eq{betau} in domain (iii), e.g.\  $N=0$ and $1<D<1.29$,
shaded regions being unstable.}
\label{flow5}
\end{figure}%
\begin{figure}[t]\centerline{
\epsfxsize=0.5\textwidth \parbox{0.5\textwidth}{\epsfbox{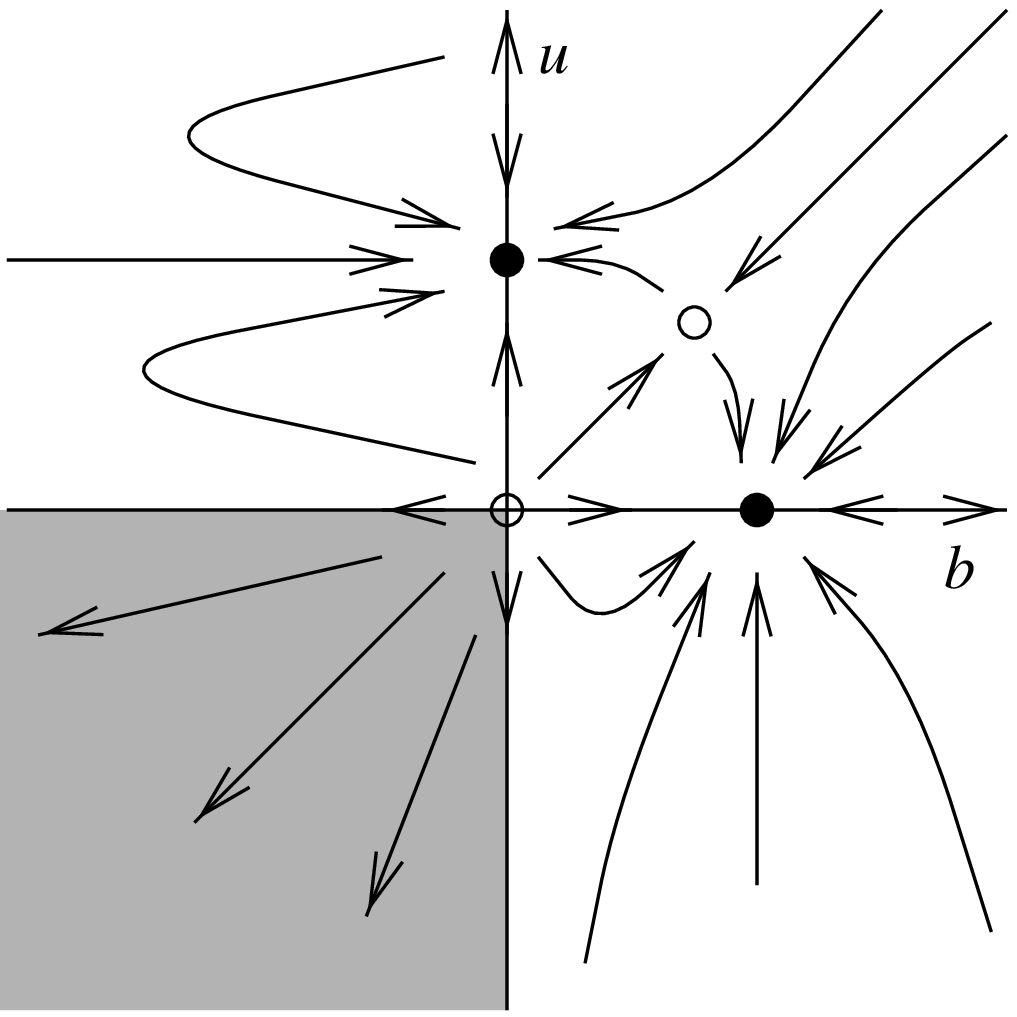}}
}\caption{RG flow from Eqs.~\eq{betab} and \eq{betau} in domain (iv), e.g.\  $N=0$ and $D$ close to 2; 
shaded regions are unstable.}
\label{flow3}
\end{figure}%
\begin{figure}[t]\centerline{
\epsfxsize=0.5\textwidth \parbox{0.5\textwidth}{\epsfbox{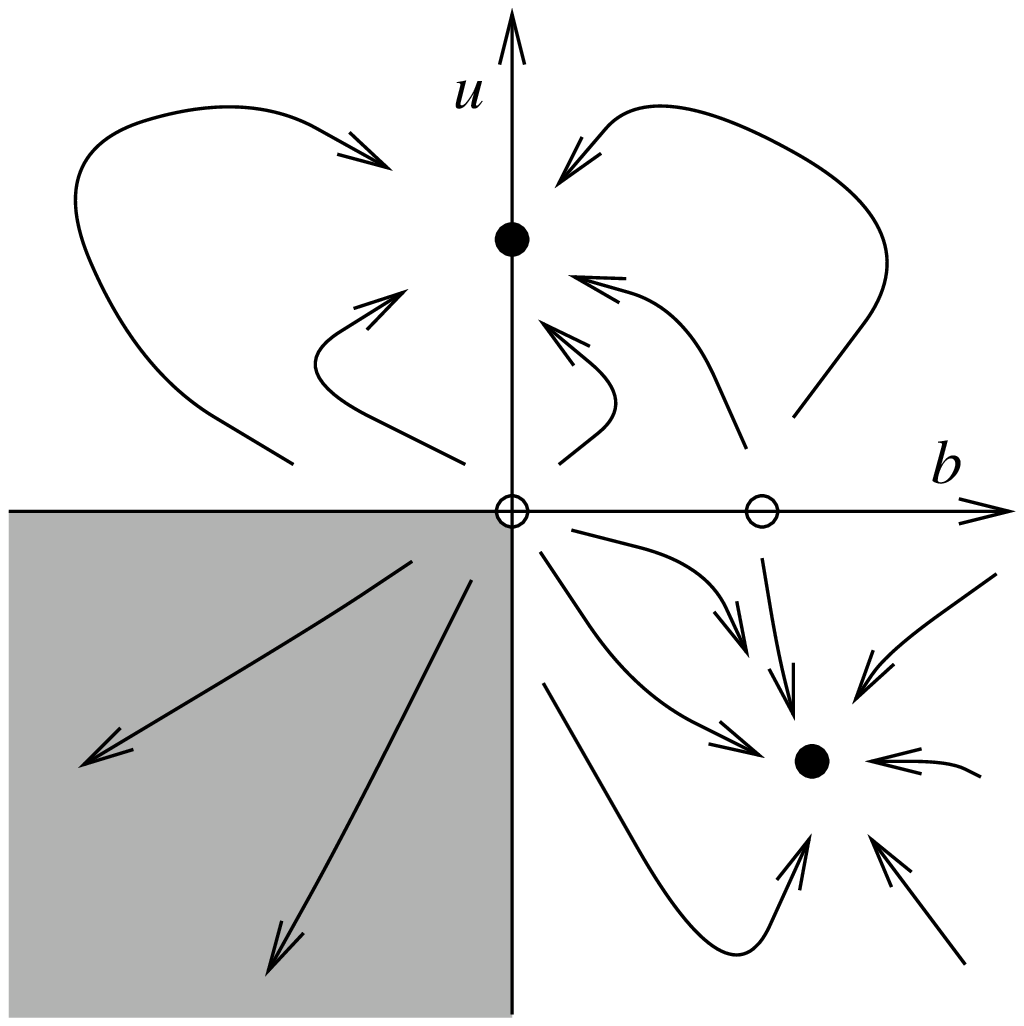}}
}\caption{RG-flow from Eqs.~\eq{betab} and \eq{betau} in domain (v), e.g.\  $N=0.2$ and $D=1.8$,
shaded regions being unstable.}
\label{flow6}
\end{figure}%
\begin{figure}[t]\centerline{
\epsfxsize=0.5\textwidth \parbox{0.5\textwidth}{\epsfbox{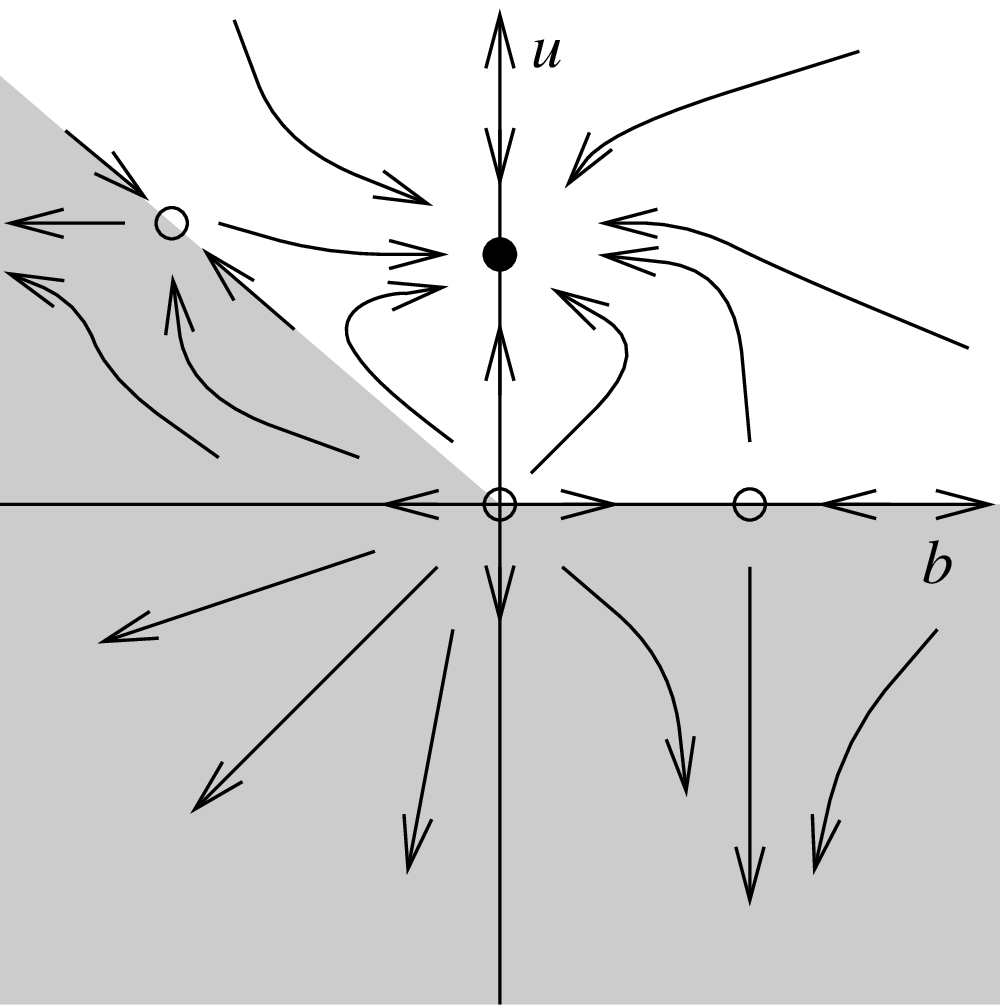}}
}\caption{RG flow from Eqs.~\eq{betab} and \eq{betau} in domain (vi), e.g.\  large $N$ and $D$ close to 2;
shaded regions are unstable.}
\label{flow4}
\end{figure}%

We may inquire as to how the above flow diagrams, with different 
stable points and stability regions,  can occur by continuously  moving around
in the $(N,D)$ plane. To demonstrate this, let us examine the sequence 
of  flow diagrams for $N=0$ and differing $D$. 
The appropriate flow diagram for $D<1$ is that of Fig.~\ref{flow1}, with the cubic
fixed point going to infinity as $D\to1$.
For $D>1$ this fixed point reappears in the upper left sector,
as in Fig.~\ref{flow5}.
Along the way, it coincides with the stable fixed point at infinity, 
and they exchange stability. 
This mechanism of changing stability from one fixed point to another 
is quite general, and occurs again when the cubic and Ising fixed 
points merge at $D\approx 1.29$. 
For $1.29<D<2$, the appropriate flow diagram is that of Fig.~\ref{flow3}.
The cubic fixed point continues to approach the Heisenberg one as
$D\to2$, resulting in very slow approaches to the fixed points in this limit.
 
We may ask whether the expanded picture presented here provides
any new insight into the behavior of tethered self-avoiding membranes 
($D=2$, $N=0$). The perturbative expansion predicts a crumpled phase
with an exponent of $\nu=0.85 <1$ in $d=3$ \cite{WieseDavid96b,DavidWiese96a}. Yet many
simulations of this 
system using models of beads and springs \cite{PlischkeBoal88,Boaletal1989,AbrEtAl89,GompperKroll92,PetscheGrest}
seem to suggest a flat phase with $\nu=1$. It is thus important to ask if
there are additions to the
standard description of such membranes in \Eq{Hamiltonian}, which can lead to a
flat configuration. The most natural candidate is a bending rigidity
$\kappa$, which is automatically generated in models of strings and beads
as pointed out in Ref.~\cite{AbrNel90}. However, it is expected that a finite $\kappa_c$
is needed in a flat phase, while the absence of a crumpled phase in simulations
may suggest $\kappa_c=0$. 
Our modified Hamiltonian with cubic anisotropy indicates that the presence
of even small $u$ for $D \to 2$ places the system within the domain
of attraction of the Ising fixed point in Fig.~\ref{flow3}.  
(The separatrix through the cubic fixed point approaches the horizontal
line as $D \to 2$.)
The exponents that we calculate for the Ising fixed point ($\nu>1$ in 
$d=3,4$  and $\nu=0.9$ in $d=5$, $\nu=0.8$ in $d=6$, and
$\nu=0.6$ in $d=8$)  are tantalizingly close to those found in the simulations of Grest \cite{Grest91}.
Yet it is hard to justify the inclusion of a finite $u$, which is meaningless
for a single membrane at $N=0$. While the presence of additional membranes
 does limit the bending of the membranes around it, the net effect
is much more than just a simple bending rigidity, as related constraints appear on
all length scales.

\section{The random bond Ising model}
\label{The random bond Ising model}
 In  this section we analyze in greater detail the model for $N=0$.
The $N\to0$ limit is interesting,  not only because of its relevance to 
self-avoiding polymers and membranes, but also for its relation to
the Ising model with bond disorder. 
To show the latter connection, we start with the field theory 
description of the random bond Ising model,  with the Hamiltonian
\be \label{random Ising}
{\cal H} = \int_r \left[\half \left( \nabla S(r)\right)^2 +\half \left(
t+\eta(r)\right) S^2(r) + u S^4(r) \right] \ ,
\ee
where $\eta(r)$ is a quenched random variable (with $\overline{\eta(r)}=0$).
Expectation values with quenched disorder can be calculated from a
partition function that is replicated $N$ times, in the limit $N\to0$
(for a review see e.g.~\cite{DotsenkoReview97}).
Averaging the replicated weight over the Gaussian random variable
$\eta(r)$, with $\overline{\eta(r)\eta(r')} = 2 \sigma \delta^d(r-r')$,
induces an interaction between different replicas with an effective Hamiltonian
\be \label{random Ising averaged}
{\cal H}_N= \int_r \sum_\alpha \left[\half \left( \nabla S_\alpha(r)\right)^2 +\frac t2 S_\alpha^2(r) + u S_\alpha^4(r) \right] - \sigma
\sum_{\alpha\beta} S_\alpha^2(r) S_\beta^2(r) \ .
\ee
The replicated system is thus controlled by a  Hamiltonian with
positive cubic anisotropy $u$, but negative $b=-\sigma$. 

A key result in the study of random bond systems is the "Harris criterion"\cite{Harris74},
which states that randomness is relevant as long as the heat capacity
exponent $\alpha$ is positive.
This is the case for the Ising model, and therefore new critical behavior
is expected for the random bond system.
In the usual field theory treatments
\cite{DotsenkoReview97,HarrisLubensky74,Khmelnitskii75,GrinsteinLuther1976},
there is no fixed point at  the 1-loop order. This  is due to the vanishing
of the denominator in \Eq{cubic FP} for $D=1$. 
However, we now have the option of searching for a stable fixed point 
by expanding about $D\ne1$.
Indeed, for  $N=0$ and $1<D<1.29$, the cubic fixed point lies in the 
upper left sector ($u>0$ and $b<0$) and is completely stable, see
 Figs.~\ref{flow5} and \ref{slowflow}. 
The extrapolation for $\nu$ at the cubic fixed point is plotted 
in Fig.~\ref{extra KIH}, where it is compared to the 
results for the Heisenberg and Ising fixed points. The divergence
of $\nu$, upon approaching $D=1$ from above, is a result of 
the cubic fixed point going to infinity as mentioned earlier. 
Upon increasing $D$, the Ising and cubic fixed points approach
each other, and merge for $D=1.29$.
 For larger values of $D$, the cubic fixed point is to the right 
of the Ising fixed point ($b_c^*>0$) and only the latter is stable. 
Given this structure, no plateau can be found for a numerical
estimate of the random bond exponent $\nu_{\mbox{\scr DO}}$, 
and we can only posit the inequality
\be \label{bound for nu}
	\nu_{\mbox{\scr DO}} > \nu _{\mbox{\scr Ising}}\ .
\ee
While \Eq{bound for nu} was derived at 1-loop order, it should 
 hold at higher orders, if no drastic subleading  corrections 
appear; since it 
merely depends on the general structure of the renormalization group
flow.

Higher loop calculations of the random bond Ising model can be 
used to expand the exponent $\nu$ in powers of $\sqrt{\E}$.
\Eq{bound for nu} can then be compared to the  2-loop 
result \cite{GrinsteinLuther1976}, which in $d=3$ reads
\be
	\nu_{\mbox{\scr DO}} = \half +\sqrt{\frac{6}{53}}\frac14\approx 0.584 \ .
\ee
This is  slightly larger  than the corresponding Ising value at 1-loop order 
($\nu\approx 0.583$), but smaller than the best known Ising result 
($\nu\approx0.631$).  
Three- and four-loop calculations were performed in Refs.
\cite{JayaprakashKatz77} and \cite{Mayer89} respectively; the latter
gives $\nu=0.6714$ and $\nu=0.6680$ depending on the resummation-method used. 
Two-loop calculations  in fixed dimension $d=3$
\cite{HolovatchShpot92} yield $\nu=0.678$.
A similar result is obtained within a modified Pad\'e-Borel
approximation of 3-loop results in Ref.\cite{JanssenOerdingSengespeick95}, 
yielding $\nu=0.666$. 
We also note that Ref.\cite{JanssenOerdingSengespeick95} stresses the
existence of very slow transients in the renormalization
group flow for small $\E$.
The extremely retarded crossover to the random bond fixed point 
may quantitatively explain the Monte-Carlo data which seem to suggest a 
disorder-dependent, and thus non-universal, behavior for the critical exponents. 
The same characteristic flow is also present at 1-loop order for our
random bond fixed point candidate,  e.g.~for $D=1.1$ as in Fig.~\ref{slowflow},
further validating our approach. 
{\begin{figure} 
\epsfxsize=0.7\textwidth \centerline{\epsfbox{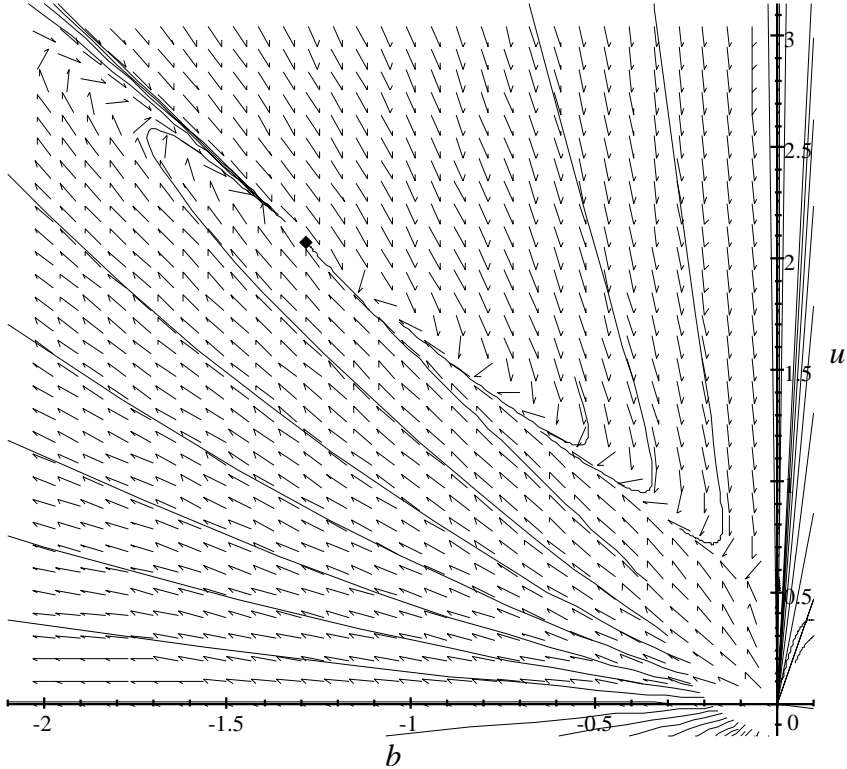}}
\caption{Renormalization group flows for $D=1.1$ and $N=0$. The cubic
fixed point is marked by a diamond}%
\label{slowflow}%
\end{figure}}%
 Within error-bars, all these results are consistent with $\alpha=0$, i.e.
the border-line value of the Harris criterion\cite{Harris74}, corresponding to
\be
\nu=\frac23\ .
\ee
There is also an exact result\cite{ChayesEtAl86} 
that the exponent $\nu$ in any random system must be greater than or equal
to $2/d$. Generalizing the ``Harris criterion" bound to manifolds yields a
limiting value of $2D/d$, which is also the mean--field value discussed earlier.

\begin{figure}[t]\centerline{
\epsfxsize=0.7\textwidth \parbox{=0.7\textwidth}{\epsfbox{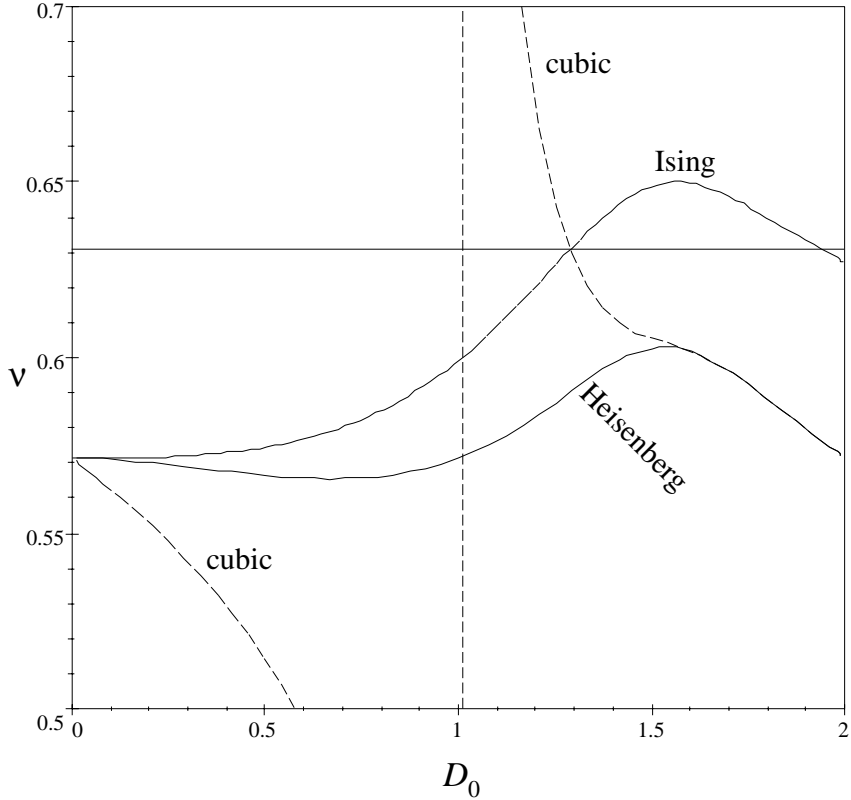}}
}\caption{Extrapolations of $\nu$ from the expansion of
$\nu d$ with $c(D)=D$, for the $O(N)$-model in three dimensions.
The values of the Heisenberg fixed point for $N=0$ are compared 
to those of the Ising and cubic fixed points. The crossing of the
latter two curves yields an estimate of
$ \nu = 0.6315 $ for the $d=3$ Ising model.}
\label{extra KIH}
\end{figure}%
The duality between the high and low temperature expansions of the
Ising model remains valid in the presence of random bonds.
As discussed above, the high temperature expansion can be presented
geometrically as a theory of loops with self-avoidance ($u>0$), and
mutual attraction ($b<0$).
We can develop a related low temperature expansion as follows:
Starting with an ordered ground state, the partition function $Z[\{J\}]$ can be
expanded as a sum over contributions of droplets of the opposite spin;
each element of the droplet surface crossing a local random bond $J$ makes
a contribution of $\exp(-2\beta J)$. 
A replicated description is obtained for $Z[\{J\}]^N$ as a sum over droplets
of $N$ different colors.
The next step is to average over all the random bonds $\{J\}$:
A specific bond may be crossed $m=0,1,\ldots,N$ times for a given term
in  $Z[\{J\}]^N$. 
Assuming that the random bonds are independently chosen from a 
Gaussian distribution of width $\sigma$, each random bond contributes a
factor of
\be
\overline{\exp(-2\beta mJ)}=\exp\left[ -2\beta \left(\overline{J}-\beta\sigma\right)m
+4\beta^2\sigma \, \frac{m(m-1)}{2}\right].
\ee
The first term in the exponent on the right hand side can be regarded as a
shift in the bond energy due to randomness. 
The second term represents a pairwise attraction between the $m$ manifolds
of different colors.
Thus the quench averaged low temperature description is of a set of 
$D=d-1$ dimensional self-avoiding droplets, with mutual attractions between
droplets of different colors. (This is easily generalized to models with
interactions defined on other elementary manifolds.)

We can now make the conjecture that the low temperature sums are
not drastically modified by restricting the droplets to tethered membranes.
This will again lead to the exponent identity in \Eq{nu hypoth}.
The extrapolations from the dual descriptions of the random bond
Ising model (at the cubic fixed point) are presented in Fig.~\ref{disorderdual}.
Because of the absence of a plateau, there are no clear points where
these curves can be compared.
The intersection of the two curves provides a specific point for extracting
exponents of the random bond model. However, as discussed in 
section \ref{Low temperature expansions of the Ising model}, at the 1-loop
level this occurs at the mean-field value of $2/3$.

\begin{figure}[t]\centerline{
\epsfxsize=0.7\textwidth \parbox{=0.7\textwidth}{\epsfbox{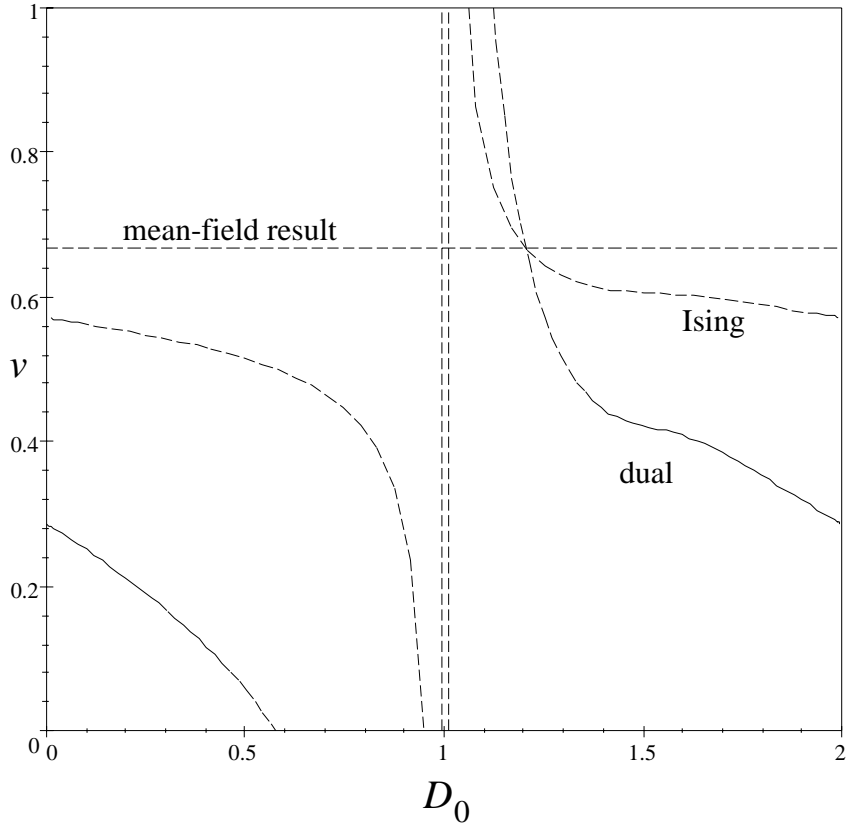}}
}\caption{Extrapolations of $\nu$ from the expansion of
$\nu d$ with $c(D)=D$, for the random bond  Ising-model in $d=3$, 
from the high and low temperature descriptions. 
Note that the intersection of the curves (from these dual descriptions) 
at the mean-field value of 2/3 is a property of the  1-loop expansion, 
as discussed in section \ref{Low temperature expansions of the Ising model}.}
\label{disorderdual}
\end{figure}%
In the previous sections we relied on a plateau in the extrapolation curves
to obtain numerical values of the exponents.
Another method for selecting a specific point of these curves is to look at
intersection points.
For example, in Fig.~\ref{extra KIH} the curves corresponding to the cubic
and Ising fixed points intersect at $D=1.29$.
Since the intersection occurs when the two fixed points coalesce on 
the $b=0$ axis, they both have Ising symmetry at this junction.
We may hope that this point yields precise exponents as all ambiguity
in the expansion point is removed.
The actual numerical prediction of
\be
	\nu^* = 0.6315 \ ,
\ee
is indistinguishable from higher-order calculations \cite{Zinn}.
However, the method is not insensitive to the extrapolation
variables used, and it is thus unclear if more precise results 
are obtained in higher order calculations.

Finally, we note that the above considerations are
easily generalized to multi-component spins subject to random bonds.
We shall denote the number of components of the field by $p$,
reserving the symbol $N$ for the number of replicas, as in the
random bond Ising model (corresponding to $p=1$).
Starting as in \Eq{random Ising} from 
\be \label{random p}
{\cal H}^p = \int_r \sum_{i=1}^p \left[\half \left( \nabla S^i(r)\right)^2 +\half \left(
t+\eta(r)\right) S^i(r)^2\right] + u \sum_{i,j=1}^p S^i(r)^2S^j(r)^2 \ ,
\ee
yields in analogy with \Eq{random Ising averaged},
after replicating and averaging over $\eta$, 
\be  \label{random p everaged}
{\cal H}^p_N= \!\int_r \sum_{\alpha=1}^N\sum_{i=1}^p \left[\half\! \left( \nabla S^i_\alpha(r)\right)^2 \!\!+\!\frac t2 S^i_\alpha(r)^2 \right] 
+u\! \sum_{\alpha=1}^N\sum_{i,j=1}^p S_\alpha^i(r)^2S_\alpha^j(r)^2 
- \sigma \!\!\sum_{\alpha,\beta=1}^N\sum_{i,j=1}^p
 S_\alpha^i(r)^2 S^j_\beta(r)^2 \ .
\ee
In the resulting $O(N)\times O(p)$-model, rotational symmetry 
is broken in the $O(N)$-sector, but remains intact in the $O(p)$-sector.
We can next develop a geometrical description of the high temperature
expansion of this field theory in the language of self-avoiding 
polymer loops, which are then generalized to membranes.
It is then easy to see that in a perturbative expansion, each 
closed loop contributes an additional factor of $p$ 
(as in the standard $O(N)$-model, where every closed loop 
is accompanied by a factor of $N$). 
All previous results are thus simply generalized by replacing $c(D)$
in the renormalization group expressions with $p\times c(D)$.
Since, according to the Harris criteorion\cite{Harris74}, non-trivial random
bond exponents are obtained in the physical dimensions of $d=2$ and
$d=3$, only for $p\leq2$, we shall not pursue this analogy further.

\section*{Acknowledgments}
It is a pleasure to thank F.\ David, H.W.\ Diehl and  
L.\ Sch\"afer for useful discussions. The major part of this 
work has been done during a visit of K.J.W.\ at MIT and he would
like to thank the MIT for its hospitality as well as the 
Deutsche Forschungsgemeinschaft for financial support through the 
Leibniz-Programm. 
The work at MIT is supported by the NSF Grant No.~DMR-93-03667.

\appendix

\section{Derivation of the RG-equations}
\label{derive RG}
In this section, we give a derivation of the renormalization group functions.
Starting from 
\be \label{appA1}
b = b_0 \,Z_b^{-1}\,Z^{-d/2}\,\mu^{-\E} \ ,
\ee
 the $\beta$-function is given through the variation of 
the renormalized coupling, at fixed values of the bare coupling and bare 
chemical potential, as
\be \label{appA2}
\beta(b):= \mu \frac{\partial }{\partial \mu}\lts_0 b \ .
\ee
From the derivative of \Eq{appA1} with respect to $\mu$, we obtain
\be
\beta(b) \left( 1+ b \frac \partial {\partial b} \ln (Z_b Z^{d/2}) \right)
=-\E b  \ ,
\ee
which solving for $\beta(b)$, yields
\be
\beta(b) =\frac{-\E b}{1+ b \frac \partial {\partial b} \ln (Z_b Z^{d/2})}
\ .
\ee

The scaling exponent $\nu$ relates the 
chemical potential $t$ and radius of gyration $R$  through
\be
R \sim t^{-\nu/D} \ .
\ee
To obtain $\nu$, we first observe that the dimensionless combination $R^2t^{\frac{2-D}D}$, 
is a function of $b$ and $t/\mu^D$ only, i.e.
\be \label{appA4}
R^2t^{\frac{2-D}D}  = f(b,t/\mu^D) \ .
\ee
Since  in addition, $t$ can be expressed as a function of $t_0$ and $b$ only, \Eq{appA4} implies
that 
\be \label{appA5}
\mu \frac{\rmd}{\rmd\mu}
\left[ R^2t^{\frac{2-D}D} \right]
= \left(\beta(b) \frac{\partial}{\partial b} - D t \frac{\partial}{\partial t}
\right) \left[ R^2t^{\frac{2-D}D} \right] \ .
\ee
Next observe that  $R_0^2 t_0^{\frac{2-D}D}$ is independent 
of the renormalization scale $\mu$. 
Replacing  bare by renormalized quantities, we obtain
 a relation for the total derivative with respect to $\mu$, as
\be 
\mu \frac{\rmd}{\rmd\mu}
\left[ R^2t^{\frac{2-D}D} Z(b) Z_t(b)^{\frac{2-D}D}
\right] =0 \ .
\ee
Combining the latter relation with \Eq{appA5} gives
\be
 \left(Dt\frac{\partial}{\partial t} -\beta(b)  \frac{\partial}{\partial b}
-\beta(b) \frac{\partial}{\partial b} \ln\left(Z Z_t^{\frac{2-D}D} \right)  \right)
\left[ R^2t^{\frac{2-D}D}  \right] =0 \ .
\ee
The scaling function of the field, describing the scaling 
of the membrane at the critical point with $\beta(b)\to0$, is thus
\be
\nu(b) = \frac{2-D}2 -\half\beta(b) \frac{\partial}{\partial b} \ln \left(Z Z_t^{\frac{2-D}D}\right) \ .
\ee
(Note that the last term cannot be dropped, as the vanishing of the 
$\beta$-function is canceled by the divergence of the derivatives
of the $Z$-factors at this point.)

\section{Reparametrization invariance}
\label{rescale}
In this appendix, we shall explore the consequences of a reparametrization
\be
x \longrightarrow x'=x \,Z_\alpha^{-1/D} \ ,
\ee
on the Hamiltonian 
\bea \label{Hamiltonian1}
{\cal H} &=& \frac Z{2-D} \int_x \half (\nabla r(x))^2 
+ b \mu^{\E}
 Z_b   
\int_x\int_y \tilde \delta^d(r(x)-r(y)) + t Z_t \Omega \ ,
\eea
for a self-avoiding membrane ($D=1$ for polymers).
(The notation for the rescaling factor anticipates renormalization factors
that we shall introduce next.)
The Hamiltonian in \Eq{Hamiltonian1} is in fact  not invariant under 
this rescaling because of the cutoff implicit in the interaction. 
In order to achieve scale invariance, the cutoff, or equivalently
the renormalization scale $\mu$, must also be rescaled to
\be
\mu \longrightarrow \mu'=\mu \,Z_\alpha^{1/D} \ .
\ee
The Hamiltonian then changes to 
\bea
{\cal H} &=& \frac {ZZ_\alpha^{\frac{2-D}D}}{2-D} 
\int_x \half (\nabla r(x))^2 
+ b \mu^{\E}
 Z_b   Z_\alpha^{\E/D-2}
\int_x\int_y \tilde \delta^d(r(x)-r(y)) + t Z_tZ_\alpha^{-1} \Omega \ .
\qquad
\eea
Comparing to the original Hamiltonian then identifies the 
new renormalization group factors
\bea 
Z'&=&Z Z_\alpha^{\frac{2-D}D} \nn\\
\label{haeh}
Z_t'&=&Z_tZ_\alpha^{-1}\\
Z_b'&=&Z_b Z_t^{-2+\E/D} \ .\nn
\eea
As discussed in the main text, see \Eqs{Rin1} and \eq{Rin2}, 
the renormalization group-functions are left unchanged by the 
transformations in \Eqs{haeh}; the most useful case being $Z_\alpha=Z_t$.

\section{Structure of the divergences and the  MOPE}
\label{Div and MOPE}
In this appendix we present a more intuitive description of 
the structure of the divergences, and the multilocal operator 
product expansion (MOPE) used  to prove renormalizability\cite{DDG3,DDG4}
and  for explicit calculations \cite{DavidWiese96a,WieseDavid96b}. 
This presentation already appears in French 
in Ref.~\cite{WiesePhD96}, but is reproduced here for 
completeness, and convenience of readers. 

We first remark that with our choice of normalizations, the free propagator 
\be
\frac1{d}\left< \half \left( r(x_1) - r(x_2) \right)^2 \right>_0 =  |x_1-x_2|^{2-D} \ ,
\ee
is the Coulomb potential in $D$ dimensions. This analogy with 
electrostatics will help us to analyze the structure of the 
divergences.
The interaction part of the Hamiltonian $\cal H$ is reminiscent
of a dipole, and can be written as 
\bea
{\cal H}_{\mbox{\scr int}} &=& 
b \int_{x_1} \int_{x_2} \tilde \delta^d(r(x_1)-r(x_2))  \nn\\
&=& b \int_{x_1} \int_{x_2} \int_k \rme^{ik ( r(x_1)-r(x_2))} \ .
\eea

The next step is to analyze the divergences appearing in the 
perturbative calculation of expectation values of observables. 
To simplify the calculations, we focus on the partition function
\be
{\cal Z}= \sum_{\mbox{\scr all states}} \rme^{-\cal H}
=\left< \rme^{-{\cal H}_{\mbox{\scr int}}} \right>_0 \ .
\ee
To exhibit the similarity to Coulomb systems, consider the second order term
\bea 
\frac{1}2 \left<  {\cal H}_{\mbox{\scr int}}^2 \right>
&=& \frac{b^2}2  
\int_{x_1} \int_{x_2}\int_{y_1} \int_{y_2} \int_k\int_p 
\left<  \rme^{ik ( r(x_1)-r(x_2))}   \rme^{ip ( r(y_1)-r(y_2))} \right>_0 \nn \\
&=& \frac{b^2}2 
\int_{x_1} \int_{x_2}\int_{y_1} \int_{y_2} \int_k\int_p \, 
\rme^{-E_{\mbox{\scr c}}} ,
\label{e:h quad}
\eea
where $E_c$ is the Coulomb-energy of a configuration of dipoles
with charges $\pm k$, and $\pm p$, respectively.
More generally, for any Gaussian measure we have
\be
	\left< \rme^{i \sum_i k_i r(x_i) } \right>_0 = \rme^{-\half \sum_{i,j} k_i k_j 
\left< (r(x_i) -r(x_j))^2\right>_0} \ .
\ee
Since for any configuration of dipoles, specified by their coordinates and charges, 
the total charge is zero, the Coulomb-energy is bounded from below, i.e.\
\be
E_{\mbox{\scr c}} \ge 0 \ . 
\ee 
This implies that
\be
 \rme^{- E_{\mbox{\scr c}} } \le 1  \ ,
\ee
and that the configurations which contribute most are those with minimal charge.

Let us apply the above observation to evaluating the integrals in \Eq{e:h quad}. 
The basic idea is to look for classes of configurations which are similar. 
The integral over the parameter which indexes such configurations is the 
product of a divergent factor, and a ``representative'' operator. 
For the case of two dipoles, one with charge $k$ and  the other with charge $p-k$
contracted together,  one only sees a simple dipole  
with charge $p$ from far away, i.e.
\bea
{\scriptstyle k \atop \scriptstyle p-k}  {\,\epsfxsize=1.5cm \parbox{1.5cm}{\epsfbox{./gm_quer.eps}}\,} {\scriptstyle -k \atop \scriptstyle -p+k} 
\, &\approx& \, {\scriptstyle p} {\,\epsfxsize=1cm \parbox{1cm}{\epsfbox{./gb.eps}}\,} {\scriptstyle p}
\ \times \ \rme^{-k^2 (|s|^{2-D} +|t|^{2-D})} \ .\nn\\
\eea
The second factor on the r.h.s.\ contains the dominant part of the 
Coulomb energy $E_{\mbox{\scr c}} = {k^2 (|s|^{2-D} +|t|^{2-D})}$ of the interaction 
between the two dipoles; $s$ and $t$ are the distances between 
the contracted ends. The integral over $k$ is now factorized, and we obtain
\be
	\int_k \rme^{-k^2 (|s|^{2-D} +|t|^{2-D})} = (|s|^{2-D} +|t|^{2-D})^{-d/2} \ .
\ee
We define the MOPE coefficient, as
\be
\bigg( \GM \bigg| \GB  \bigg) = \left(|s|^{2-D} +|t|^{2-D}\right)^{-d/2}\ .
\ee
The MOPE therefore gives a convenient and powerful tool to calculate the
dominant and all subdominant contributions from singular configurations.

 For the sake of  completeness, let us still calculate the two other MOPE coefficients used 
in the text. The first is for small $|x-y|$, ($\nu_0=\frac{2-D}2$)
\bea
 \vphantom{\hbox{\GH}}_x\GH_y &=&
\int_k \,:\rme^{i k r(x)}:\, :\rme^{-ikr(y)}: \nn \\
&=&  
\int_k \,:\rme^{i k (r(x) -r(y))}: \rme^{-k^2 |x-y|^{2\nu_0}} \nn \\
&=& \int_k \left[ \mbox{\bf 1} - \frac{k^2}{2d} \,:\left[r(x)-r(y)\right]^2:\, + \ldots \right]
 \rme^{-k^2 |x-y|^{2\nu_0}} \nn \\
&=& |x-y|^{-\nu_0 d} \,\mbox{\bf 1} - \frac14 \,:\left[ (x-y) \nabla r\left(\frac{x+y}{2}\right) \right]^2 :\, |x-y|^{-\nu_0(d+2)} + \ldots
\nn \\
&=& |x-y|^{-\nu_0 d} \,\mbox{\bf 1}\, - \frac 1 {2D}  |x-y|^{D-\nu_0 d}
\,\GO\,+ \ldots
\ .
\eea
(The normal ordered operator $:\!\!O\!\!:$ indicates that all self-contractions of $O$
have been subtracted.)
We also have to show the factorization property of $\FD$, which is the product
of two $\tilde \delta^d$-interactions
\be \label{uiui}
 \FD
\raisebox{-5.5\pmm}[0mm][0mm]{\makebox[0mm][l]{$\hspace{-18\pmm}\scriptstyle u\ \  \
x\, y\, z$} 
}
=\int_{k,p} \,:\rme^{i k r(u)}:\, :\rme^{-ikr(x)}:
 \,:\rme^{i p r(y)}:\, :\rme^{-ipr(z)}:
\ .
\ee
We want to study the contraction of $x$, $y$, and $z$, and look for all 
contributions which are proportional to 
\be
\GB= \int_{k} \,:\rme^{i k r(u)}:\, :\rme^{-ikr((x+y+z)/3)}: \ .
\ee
The key-observation is that in \Eq{uiui} no contraction with $:\rme^{-ikr(x)}:$
contributes. All such contractions yield a factor of $k$, which  after integration 
over $k$ results in  derivatives of the $\tilde \delta^d$-distribution. This is equivalent 
to stating that as long as contributions proportional to $\GB$ are studied,
the following factorization property holds:
\be
	\FD = \GB \times \GH
\ .
\ee
This is the reason why, in the massless scheme, divergences proportional to   $\FD$
are already eliminated through a counter-term for $\GH$, i.e.\ by  the
renormalization factor $Z$.

\section{Renormalization for infinite membranes}
\label{Ren for inf mem}
In this appendix, we give a short summary of the renormalization procedure
for infinite membranes, and derive the one-loop counter-terms. 
This is a simplified version of the corresponding section in 
Ref.~\cite{WieseDavid96b}, which is included here again
mainly for completeness.

Let us start with a single dipole:
When its end-points $(x,y)$  are contracted towards  a point
(taken here to be the center-of-mass $z=(x+y)/2$), the MOPE is
\begin{equation}
\label{e:mopecoef}
 \vphantom{\hbox{\GH}}_x\GH_y 
=
\bigg( \vphantom{\hbox{\GH}}_x\GH_y \bigg|\, \mbox{\bf 1} \bigg) \mbox{\bf 1} \ +\ 
\bigg( \vphantom{\hbox{\GH}}_x\GH_y \bigg| \GO \bigg)
\GO  \ + \ldots \ . \nn  
\end{equation}
The first MOPE-coefficients are given explicitly by
\bea
\bigg( \vphantom{\hbox{\GH}}_x\GH_y \bigg|\, \mbox{\bf 1}\bigg) &=&|x-y|^{-\nu_0 d} \ ,
\nn \\
\bigg( \vphantom{\hbox{\GH}}_x\GH_y \bigg| \GO  \bigg)&=&
 -\frac1{2D} |x-y|^{D-\nu_0 d} \ ,
\eea
and where $\GO$ denotes the local  operator
\begin{equation}
\GO =\half (\nabla r(x))^2  \ .
\end{equation}
The integral over the relative distance $x-y$ for
 $\bigg( \vphantom{\hbox{\GH}}_x\GH_y \bigg| \GO\bigg) $ is,
at $\E=0$, logarithmically divergent. 

The simplest contraction resulting in a dipole is when two dipoles coalesce.
The corresponding MOPE-coefficient is
\be
\bigg( \GM \bigg| \GB \bigg) 
= \left( |x|^{2\nu_0} + |y|^{2\nu_0} \right)^{-d/2} \ ,
\ee
where $x$ and $y$ are now the relative distances inside the two subsets.
Another possibility is to consider the contraction $\FD$.
But as we have shown in appendix \ref{Div and MOPE},
this does not give a new divergence.

In the next step, counter-terms are introduced to subtract these divergences.
We have to distinguish between counter-terms for 
relevant operators and those for marginal operators.
The former can be defined by analytic continuation, while the latter require
a subtraction scale. 
Indeed, the divergence for $\mbox{\bf 1}$ is given by the integral
\bea
	\int_{\Lambda^{-1}< |x-y| < L} \bigg( \vphantom{\hbox{\GH}}_x\GH_y \bigg| \mbox{\bf 1}\bigg)
	&=&\int_{\Lambda^{-1}}^{L} \frac{dx}{x} x^{D-\nu_0 d} \\
	&=&\frac1{D-\E} \left( \Lambda^{D-\E} -L^{\E-D}\right) \ ,\nn
\eea
where $\Lambda$ is a high-momentum UV-regulator, and $L$ a large distance
regulator.
For $\E \approx 0$, this is UV-divergent but IR-convergent.
The simplest way to subtract this divergence is therefore to replace the
dipole operator by 
\be
\strut_x\GB\strut_y\  \longrightarrow\  \strut_x\GB\strut_y\,-\,
\strut_x\GBdotted\strut_y \ ,
\ee
where $\strut_x\GBdotted\strut_y = |x-y|^{-\nu_0 d} $. 
This amounts to adding to the bare Hamiltonian, the UV-divergent
counter-term
\be
\Delta{\cal H}[r] = -{b\over 2}\,	\int_x \int_y |x-y|^{-\nu_0 d} \ , 
\ee
which is a pure number, and thus does not change the expectation value of any
physical observable.

We next consider marginal operators:
In the MOPE of \Eq{e:mopecoef},
the integral over the relative distance of 
${\displaystyle\int_{x-y}}\bigg( \vphantom{\hbox{\GH}}_x\GH_y \bigg|
\GO \bigg) \GO $
is logarithmically divergent at $\E=0$.
In order to find the appropriate counter-term, we use dimensional
regularization, i.e.\ set $\E>0$.
An IR-cutoff $L$, or equivalently a subtraction momentum scale $\mu=L^{-1}$,
has to be introduced in order to define the subtraction operation.
As a general rule, let us integrate over all distances appearing in the
MOPE-coefficient, bounded by the subtraction scale $L=\mu^{-1}$, giving
\be \label{e:MOPEL}
\bigg< \GH \bigg| \GO \bigg>_L:=
\int_{|x-y|<L} \bigg( {}_x\GH_y \bigg| \GO \bigg)  = L^\E\,f(\E,D)
\ .
\ee
 Following Refs.~\cite{DDG3,DDG4,WieseDavid96b,DavidWiese96a,WieseDavid95},
 we use a minimal subtraction scheme (MS).
The internal dimension $D$ of the membrane  is kept fixed, and 
\Eq{e:MOPEL} is expanded as a Laurent series in $\varepsilon$, which here
starts at $\varepsilon^{-1}$.
Denoting by $\Res \big<\ \big|\ \big>$, the residue of the term of order
$\varepsilon^{-1}$ of the Laurent expansion of $\big<\ \big|\ \big>_L$ for
$L=1$, the residue of the pole in  \Eq{e:MOPEL} is found to be
\be
\Res \bigg< \GH \bigg| \GO \bigg> = -\frac{1}{2D} 
\ .
\ee
It is this pole that is  subtracted in the MS-scheme by adding to the 
Hamiltonian a counter-term 
\begin{equation}
\Delta{\cal H}[r] =- \frac b\E \, \Res \bigg< \GH \bigg| \GO \bigg>  \,
\int_x \GO_x \ .
\end{equation}

Similarly, the divergence arising from the contraction of two dipoles into 
a single dipole is subtracted by a counter-term proportional to the residue of the
single pole of
\bea
\bigg< \GM \bigg| \GB \bigg>_L &=&
	\int_{|x|<L}\int_{|y|<L} \bigg( \GM \bigg| \GB \bigg) \nn\\
&=&\int_{|x|<L}\int_{|y|<L}\left( |x|^{2\nu_0} + |y|^{2\nu_0} \right)^{-d/2}
\ . \nn\\
\eea
Integrating the above yields
\be
\Res \bigg< \GM \bigg| \GB \bigg> = \frac1{2-D} \frac{\Gamma\left(\frac D{2-D}\right)^{2}}
{\Gamma\left(\frac{2D}{2-D}\right)}
\ .
\ee
As a result,  the model is UV-finite if we use the
renormalized Hamiltonian 
\bea \label{e:Ham1}
{\cal H}_{R}[ r]&=& \frac{Z}{2-D}\int_x \half 
\big(\nabla r(x)\big)^2 + b Z_b \mu^\varepsilon \int_x\!\int_y\!
\tilde \delta^d\big ( r(x)-  r(y)\big ) \ ,
\eea
instead of the bare Hamiltonian ${\cal H}[ r]$.
Here, $r$ and $b$ are the renormalized field and coupling constant,
and $\mu=L^{-1}$ is the renormalization momentum scale,
and the renormalization factors at one loop order are
\bea 
Z&=& 1-(2-D) \Res \bigg< \GH \bigg| \GO \bigg> \frac{b}\E  \ ,
\label{e:Z1}
\\
Z_b&=&1+ \Res \bigg< \GM \bigg| \GB \bigg>\frac b \E \ .
\label{e:Zb1} 
\end{eqnarray}

The renormalized field and coupling constants are  re-expressed in terms of their bare 
counterparts  through \begin{equation}
 r_0(x) = Z^{1/2}\, r(x)
\ ,\qquad
b_0 = b\,Z_b\,Z^{d/2}\,\mu^\E \ .
\end{equation}
 Following the analysis of Refs.~\cite{DDG3,DDG4,WieseDavid96b,DavidWiese96a},
the renormalization group $\beta$-function and $\nu$ (the anomalous scaling
dimension  of $ r$) are obtained from the variation of the
coupling constant and the field with respect to the renormalization scale $\mu$,
keeping the bare couplings fixed.
They are written in terms of $Z$ and $Z_b$ as
\bea
\label{e:beta}
\beta(b) &=&
\mu \frac{\partial}{\partial \mu }\lts_{b_0} b
=\frac{-\E b} {1+ b\frac{\partial}{\partial b } \ln Z_b +
\frac{d}2 b \frac{\partial}{\partial b} \ln Z}\ ,\\
\label{e:nu}
\nu (b) &=&
\frac{2-D}{2}-\half \mu \frac{\partial}{\partial \mu }\lts_{b_0} \ln Z \nn\\
&=&
\frac{2-D}2 -\half \beta(b) \frac{\partial}{\partial b} \ln Z \ .
\end{eqnarray}

\section{Other topologies}
\label{Other topologies}
Throughout the manuscript we have generalized polymer loops at $D=1$
to closed hyperspheres at $D\neq 1$.
Natural questions are whether other topologies may also be used,
or if it is possible  to sum over all topologies with appropriate weights. 
%
The only equation where the topology enters, is that of the partition function
for a single polymer or membrane in \Eq{Z1(0)mem}. Following 
David et al.\ \cite{DDG4}, this is generalized to 
\bea 
{\cal Z}_{1,\chi}^{(0)} &=& \frac{c(D)}{D} \int \frac{\rmd \Omega} \Omega\, \Omega^{1+{\chi d}/{6D} -\nu_0 d/D} \rme^{-t \Omega} \nn\\
&=& \frac{c(D)}D \Gamma\left( \frac\E D +\frac{\chi d}{6D}-1 \right) t^{\E/D+\frac{\chi d}{6D}-1} \ ,
\label{Z1(0)mem gen top}
\eea
where $\chi$ is the Euler characteristics of the membrane; $\chi=0$ corresponds
to a sphere, $\chi=1$ to a 1-torus, and so on.
Note that this equation is strictly correct
only  for $D=2$, but that we make an analytic continuation for 
arbitrary $D$ in order to keep the anomalous contribution from the 
topology. 
(Topological anomalies, related to trace anomalies
in conformal field theory \cite{PhilippeCFT}, occur only in integer dimensions.)
Expanding now for $\E=0$, the  1-torus gives an 
additional contribution at $D=4/3$, and $d=8$. 
Thus for $D>4/3$, the torus is irrelevant, higher topologies are even 
more irrelevant, and their neglect is justified. 
In principle, for $D<4/3$, we can perform a double $\E$-expansion about this point. 
The second expansion parameter is 
\be
\delta=\E  +\frac{ d}{6}-D
\ .
\ee
We can then introduce four different couplings. One coupling $b$ on
the same object (be it a torus or a sphere), a second coupling $g$
between spheres, a third coupling $u$ between spheres and tori,
and a fourth coupling between tori. 
We leave such calculations for the future.

\bibliography{../citation/citation}

\bibliographystyle{./KAY}

\end{document}